%% file: main.tex
\documentclass[sigconf]{acmart}

\setcopyright{none}             
\renewcommand{\footnotetextcopyrightpermission}[1]{} 

\usepackage{url}

\usepackage{tikz}
\usepackage{amsmath}

\usepackage{xcolor}

\usepackage{cleveref}
\usepackage{enumitem}

\usepackage{graphicx}
\usepackage{xspace}
\usepackage{subfigure}
\usepackage{cleveref}
\usepackage{booktabs}
\usepackage{multirow}
\usepackage{newunicodechar}
\usepackage{pifont} 
\usepackage{float}
\usepackage{threeparttable} 
\usepackage{enumitem} 
\usepackage{balance}

\newunicodechar{−}{\textminus}
\crefname{figure}{Fig.}{Figs.}
\crefname{table}{Table}{Tables}
\crefname{section}{Sec.}{Secs.}
\crefname{equation}{Eq.}{Eqs.}

\usepackage[dvipsnames]{xcolor}

\newcommand{\xxx}{\texttt{GhostTac}\xspace}

\usepackage{hyperref}

\begin{document}

\title[GhostTac: Manipulating Tactile Sensors without Physical Contact]
{\xxx: Manipulating Tactile Sensors without Physical Contact}

\settopmatter{authorsperrow=3}

\author{Kun Wang}
\authornote{Both authors are co-first authors.}
\affiliation{%
  \institution{Zhejiang University}
  \city{Hangzhou}
  \country{China}
}
\email{eewk@zju.edu.cn}

\author{Xuancun Lu}
\authornotemark[1]
\affiliation{%
  \institution{Zhejiang University}
  \city{Hangzhou}
  \country{China}
}
\email{xuancun\_lu@zju.edu.cn}

\author{Ruochen Zhou}
\authornote{Ruochen Zhou is the corresponding author.}
\affiliation{%
  \institution{Hong Kong University of Science and Technology}
  \city{Hong Kong}
  \country{China}
}
\email{zrccc@ust.hk}

\author{Kai Wang}
\affiliation{%
  \institution{Zhejiang University}
  \city{Hangzhou}
  \country{China}
}
\email{eekaiwang@zju.edu.cn}

\author{Tongjun Ye}
\affiliation{%
  \institution{Zhejiang University}
  \city{Hangzhou}
  \country{China}
}
\email{yetongjun@zju.edu.cn}

\author{Yihao Shao}
\affiliation{%
  \institution{Zhejiang University}
  \city{Hangzhou}
  \country{China}
}
\email{yihaoshao@zju.edu.cn}

\author{Chen Yan}
\affiliation{%
  \institution{Zhejiang University}
  \city{Hangzhou}
  \country{China}
}
\email{yanchen@zju.edu.cn}

\author{Xiaoyu Ji}
\affiliation{%
  \institution{Zhejiang University}
  \city{Hangzhou}
  \country{China}
}
\email{xji@zju.edu.cn}

\author{Wenyuan Xu}
\affiliation{%
  \institution{Zhejiang University}
  \city{Hangzhou}
  \country{China}
}
\email{wyxu@zju.edu.cn}

\renewcommand{\shortauthors}{Kun Wang et al.}

\input{sections/abstract}

\maketitle
\pagestyle{plain}     
\thispagestyle{plain} 

\input{sections/introduction}

\input{sections/background}

\input{sections/threatmodel}

\input{sections/preliminary_analysis}

\input{sections/design}

\input{sections/Evaluation}

\input{sections/discussion}

\input{sections/related_work}

\input{sections/conclusion}

\bibliographystyle{ACM-Reference-Format}
\bibliography{reference/reference.bib}
\appendix

\input{sections/appendix}

\end{document}

%% file: sections/abstract.tex
\begin{abstract}

Tactile sensors are integral components of modern robotic systems, enabling robots to perceive and interact with the physical environment through tactile feedback. 
Despite their importance, the physical-layer security of tactile sensors has received little attention in prior work.
In this paper, we present \xxx, to the best of our knowledge, the first contactless attack that manipulates tactile sensing via electromagnetic interference (EMI).
We identify that EMI exploits the nonlinear rectification and limited bandwidth amplification effects, allowing carefully crafted EMI signals to be converted into a persistent DC offset that bypasses on-board filtering and induces stable measurement deviations.
Building on this mechanism, \xxx enables fine-grained and controllable manipulation of sensor outputs by reshaping the spatial distribution and manipulating the magnitude at the targeted location.
Such interference can induce unintended and harmful robot behaviors, such as causing a domestic robot to exert excessive force, resulting in physical damage or human injury.
We evaluate \xxx on 10 sensor modules and 2 dexterous hands, covering 15 tactile sensors of different types, and demonstrate consistent attack effectiveness across all tested devices. We further present three case studies on tactile grasping, slip detection, and material classification to illustrate practical impacts in real robotic tasks.
We envision that our findings shed light on a new physical attack vector against tactile sensing in robotic systems.

\end{abstract}

%% file: sections/introduction.tex
\section{Introduction}

The rapid development of robotic systems operating in unstructured environments has greatly expanded the scope of autonomous interaction~\cite{duan2022survey,ma2024survey}.
At the core of this capability lies tactile sensing, which enables robots to perceive contact forces, infer material properties, and adapt their actions with human-like dexterity.
Tactile feedback underpins a wide range of critical robotic functions, including stable grasping, precise assembly, and safe physical interaction~\cite{cui2020grasp,cui2021toward,yang2025bitla}.
Reflecting its growing adoption, the global tactile sensor market reached USD 16.4 billion in 2024~\cite{tactile-sensor-market}.
Despite this central role, the \textit{physical-layer security} of tactile sensing remains unexplored, raising concerns about the reliability and safety of tactile-driven robotic systems, particularly in open and physically accessible environments.

\begin{figure}[t]
    \centering
    \includegraphics[width=0.9\linewidth]{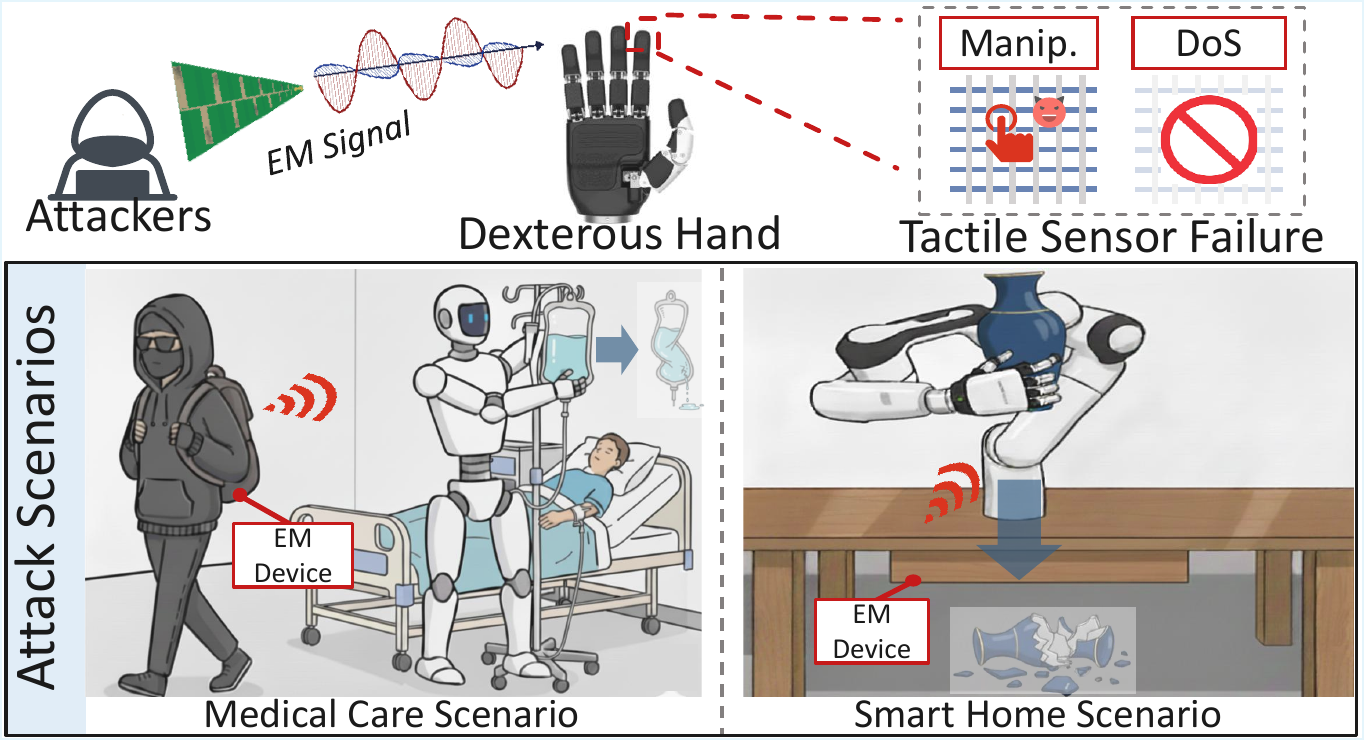}
    \caption{An illustration of \xxx attack.  
    Attackers can emit pre-designed EM signals to manipulate sensor readings or induce a DoS condition. Such interference disrupts tactile sensor measurements remotely, causing the dexterous hand to damage delicate medical devices or drop an antique vase. }
    \label{fig1}
\end{figure}

In this paper, we present \xxx, a new EMI-based attack that enables both contactless, fine-grained manipulation of tactile sensor measurements and denial-of-service (DoS) induction. 
By corrupting tactile feedback, \xxx can mislead robotic systems into risky behaviors. 
As illustrated in~\cref{fig1}, a robot may exert excessive force and damage delicate equipment such as intravenous infusion devices used in emergency care, or apply insufficient force and drop fragile objects
\footnote{Video demos are available at \url{https://ghosttac.github.io/GhostTacCCS.io/}}.
Notably, \xxx requires no physical contact, allowing attackers to stealthily launch the attack using portable EM devices carried by passersby or via pre-deployed hardware.

To achieve \xxx, we need to address the following two challenges:

\begin{figure*}[h]
    \centering
    \includegraphics[width=0.99\linewidth]{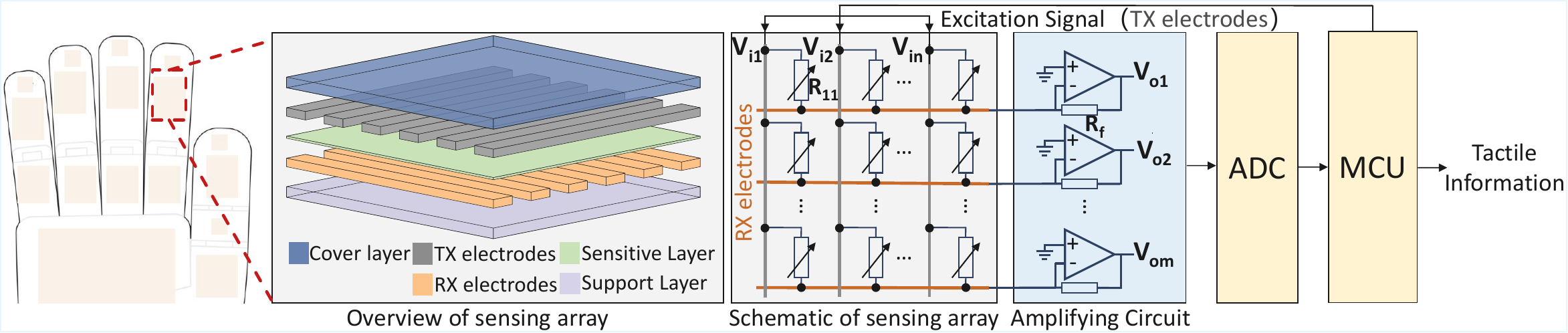}
    \caption{An overview of the tactile sensor with its corresponding system architecture. 
    The sensor converts applied pressure into a change in resistance or capacitance, which the sensing array measures as a voltage signal. This signal is subsequently amplified, digitized by an ADC, and processed by an MCU to produce accurate tactile information.}
    \label{fig-back-stru}
\end{figure*}

(1) \textit{How to effectively inject attack signals into tactile sensors?}
Contemporary electronic systems are designed to resist electromagnetic interference through electromagnetic compatibility (EMC) testing~\cite{williams2016emc} and extensive mitigation techniques, including shielding~\cite{geetha2009emi} and layout optimization~\cite{montrose2004emc}.
To bypass these defenses, we conduct a systematic analysis of tactile sensor architectures using datasheets and manuals to identify components susceptible to EMI coupling.
We then perform localized frequency sweeps using near-field probes to map coupling points, revealing that both the sensing array and communication channels are susceptible to EMI coupling.
However, direct high-frequency interference is ineffective because on-board filters attenuate it, and high-frequency AC interference cannot induce stable deviations in DC tactile measurements.
To overcome this limitation, we exploit transistor-level nonlinearities and bandwidth-limited amplification in the sensor’s operational amplifiers, which convert injected high-frequency EMI into a persistent DC offset that bypasses filtering and directly distorts sensor measurements. 
In addition, EMI injected into communication channels disrupts data transmission, triggering system freezes or protection mechanisms that render the sensor non-functional.

(2) \textit{How to create a controllable interference?}
While broad EMI can disturb sensor readings, practical attacks require fine-grained control over \textit{where}, \textit{when}, and \textit{how strongly} interference is injected, to fabricate realistic tactile events that deceive robotic operation. Achieving this level of control is challenging due to the proprietary and complex internal circuitry of commercial tactile sensors.
We address this challenge by exploiting the sequential scanning architecture of tactile arrays. 
By synchronizing malicious signals with the sensor’s excitation cycle, \xxx injects interference only when selected electrodes are active. Through parameterized modulation of baseband signals, including delay, duty cycle, amplitude, and period, \xxx realizes fine-grained and controllable interference injection, enabling real-time manipulation over force magnitude (both positive and negative interference), injection location, interference area width, and dynamic contact patterns.

We evaluate \xxx on commercial off-the-shelf sensor modules and 2 commercial dexterous hands, covering 15 tactile sensors from 8 manufacturers.
Across the entire test set, all evaluated sensors exhibit vulnerability to EMI-based attacks.
We achieve 100\% success in controllable manipulation of force magnitude, injection location, interference width, and dynamic contact patterns in laboratory settings.
To demonstrate real-world impact, we present system-level case studies on robotic grasping, slip detection, and material classification, showing that \xxx can reliably induce unintended and harmful behaviors, including damaging fragile objects and causing grasp failure.

Our contributions are summarized as follows:
\begin{itemize}
\item We identify a common physical-layer vulnerability in tactile sensors, where EMI induces persistent DC offsets via nonlinear rectification and limited-bandwidth amplification, thereby manipulating sensor outputs.
\item We design \xxx, the first contactless and controllable EMI attack on tactile sensors that enables fine-grained manipulation of force, location, width, and dynamic patterns.
\item We evaluate \xxx on 15 commercial tactile sensors from 8 manufacturers, achieving 100\% success across all types of manipulation, with demonstrated real-world impact on grasping, slip detection, and material classification.
\end{itemize}

%% file: sections/background.tex
\section{Background}\label{background}

\subsection{Tactile Sensors Overview}
Tactile sensors are input devices, typically mounted on dexterous hands or grippers, used to detect pressure changes and facilitate grasping tasks. They function by transducing pressure magnitude and distribution into electrical signals.
The sensor array dominates the market as it can detect pressure distribution through multiple sensing units. 
Commercial devices may be equipped with multiple tactile sensors to enhance user interaction and functionality. This sensor structure typically consists of a cover layer, RX electrodes (RXs), a sensitive layer, TX electrodes (TXs), and a support layer~\cite{zou2017novel,sundaram2019learning}. An overview of the tactile sensor, along with its corresponding system architecture, is shown in~\cref{fig-back-stru}.
In addition, tactile sensors employ diverse transduction mechanisms, including piezoresistive, capacitive, and piezoelectric. Despite these distinct signal generation physics, most devices share a fundamental architecture characterized by TXs and RXs~\cite{luo2025tactile,wang2015recent}.

\subsection{Tactile Sensor Principles}

\subsubsection{System Architecture}
As illustrated in~\cref{fig-back-stru}, this system architecture typically comprises four main parts: a sensing array, amplifying circuits, an analog-to-digital converter (ADC), and a microcontroller unit (MCU)~\cite{zhu2022recent}.  The sensing array is an $m \times n$ grid formed by $m$ RXs and $n$ TXs.

The signal acquisition and processing follow a sequential path. The MCU begins by implementing the scan driving method, applying excitation signals sequentially to the TXs. Contact pressure alters the resistance at the intersections of the TX and RX grids. 
An amplifying circuit based on operational amplifiers (op-amps) acquires the signals from the RXs, converting the weak inputs induced by resistance variations into amplified analog voltages. 
This circuit provides high-impedance input and low-impedance output to minimize noise and enhance stability. Subsequently, the ADC digitizes these voltages and sends the data to the MCU. Finally, the MCU processes this digital information to extract precise tactile data, including force distribution and values for each sensing unit.

\begin{figure}[t]
    \centering
    \includegraphics[width=1.0\linewidth]{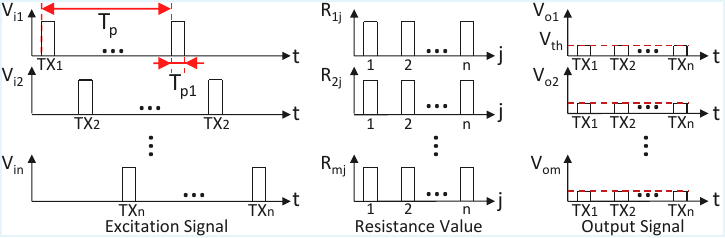}
    \caption{Waveform diagrams of the scan driving method. The excitation signals ($V_{i1}$, $V_{i2}$, …, $V_{in}$) are sequentially applied to the $n$ TXs. For each active input, the output voltages ($V_{o1}$, $V_{o2}$, …, $V_{om}$) are simultaneously read from the $m$ op-amps outputs associated with the $m$ RXs.
    }
    \label{fig-back-scan}
\end{figure}
\subsubsection{Resistive Sensing}
Resistive sensing is the method to measure resistance changes at an individual sensing unit. The core principle is that when contact pressure is applied to a unit (at the intersection of a TX and RX electrode), its material properties change, decreasing its resistance ($R_{11}$). To measure this, an excitation signal ($V_{i1}$) is applied to the TX electrode and processed by an amplifier circuit. Consequently, the output of the amplifier ($V_{o1}$) is calculated as:
\begin{equation}
    V_{o1}= - \frac{R_f}{R_{11}}V_{i1}
\end{equation}

When contact pressure increases, $R_{11}$ decreases, thereby increasing the magnitude of the output voltage. The MCU then processes this voltage to calculate the precise pressure applied to that unit.

\subsubsection{Scan Driving Methods}
A scan driving method (SDM) is used in combination with resistive sensing to measure the pressure across all tactile units in the array \cite{wang2022ghosttouch}. A widely adopted technique is the time-interleaved sensing method, known for its high processing rate. This SDM works by sequentially exciting the TXs one at a time, as illustrated by the waveforms in \cref{fig-back-scan}. 
Since only one TX electrode is active at a time, the system can pinpoint the contact location by identifying the specific TX-RX intersection.

%% file: sections/threatmodel.tex
\section{Threat Model}

Along with other EMI attack works~\cite{yang2024rethink,jiang2022wight,jiang2024ghosttype,jin2024phantomlidar}, we make the following assumptions about \xxx: 
\begin{itemize}
    \item \textbf{Attack goal:} The attacker’s goal is to manipulate or DoS the target tactile sensor measurements in a contactless manner, causing grasping failures or material misclassification.
    \item \textbf{Victim device:} The victim is a device equipped with tactile sensors requiring precise control, such as a robotic arm used for domestic services or patient care.
    \item \textbf{Prior knowledge:} The prevalence of COTS hardware allows attackers to obtain a device identical to the victim's model. Consequently, they can reverse engineer the necessary parameters for signal optimization.
    
    \item \textbf{Attacker’s capability:} We assume the attacker can approach the target device to emit EM signals but cannot physically touch or tamper with it to avoid detection. Specifically, the attacker may pre-place a malicious EM device near the target to trigger it from a distance or blend into a crowd as a passerby to initiate the attack surreptitiously, thereby ensuring both stealth and severe consequences.

\end{itemize}

%% file: sections/preliminary_analysis.tex
\section{Principle of \xxx Attack}\label{sec-principle}
In this section, we perform a systematic security analysis of tactile sensors. We first detail the underlying principles of EMI coupling and conduct the feasibility experiment. Next, we explain how high-frequency signals interfere with measurements or cause DoS. Finally, we validate these mechanisms through physical experiments.
\subsection{EMI Coupling} 
\subsubsection{Principle Analysis}
The EMI coupling mechanism is governed by Maxwell's equations. Specifically, Faraday's law of induction states that a time-varying magnetic field ($\mathbf{B}$) passing through a surface ($\Sigma$) induces an electric field ($\mathbf{E}$) along its closed boundary loop ($\partial \Sigma$):
$\oint_{\partial \Sigma} \mathbf{E} \cdot \mathrm{d} \mathbf{l} = -\frac{d}{dt} \int_{\Sigma} \mathbf{B} \cdot \mathrm{d} \mathbf{A}$.

For tactile sensors, the measuring circuit forms the loop ($\partial \Sigma$), enclosing an effective area ($\Sigma$).  An external EM signal introducing a magnetic field through this surface induces a voltage in the circuit. This voltage superimposes on the original signal, distorting the resistance readings and leading to inaccurate measurements.

As a first-order approximation in the far-field regime, the power
received by the victim circuit is modeled using the Friis transmission
equation:
\begin{equation}
    P_{i}=G_{t} G_{r}\left(\frac{\lambda}{4 \pi D}\right)^{2} P_{t}
\end{equation}
where $G_t$ and $G_r$ are the transmitter and receiver antenna gains, $\lambda$ is the EMI signal's wavelength, and $D$ is the distance between the antenna and the victim circuit.

\begin{figure}[t]
    \centering
    \includegraphics[width=1.0\linewidth]{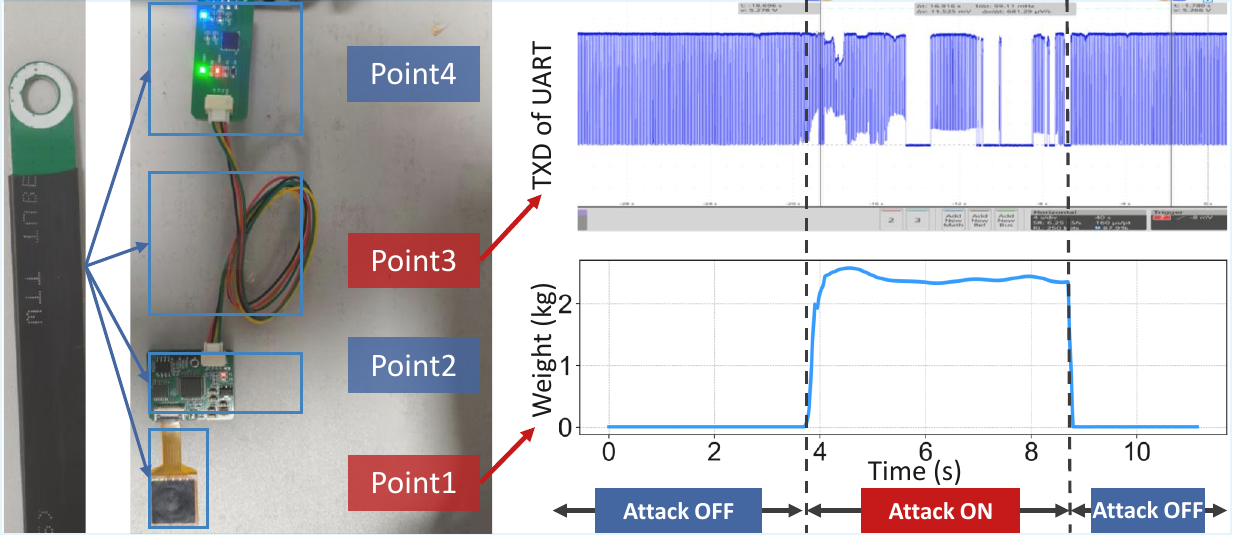}
    \caption{Illustration of the four EMI injection points and resulting signal, which depict effective injection results: induced static deviation in weight measurements (Point 1) and signal integrity loss on the TXD signal (Point 3).}
    \label{fig-pre-inj}
\end{figure}
\begin{figure}[t]
    \centering
    \includegraphics[width=1.0\linewidth]{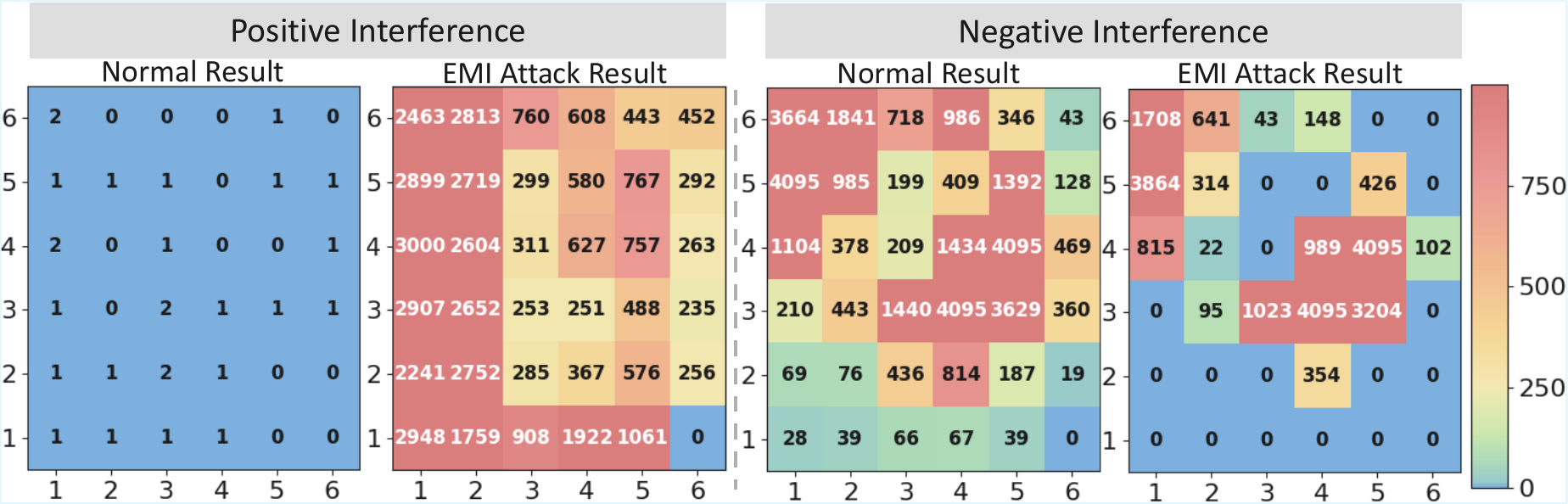}
    \caption{A visual illustration of measurement variation in tactile sensors showing that positive interference increases the measurement, while negative interference decreases it.}
    \label{fig-pre-experi}
\end{figure}

\subsubsection{Experiment Validation}

To identify the vulnerable component of a tactile sensor~\cite{legact}, we used a magnetic probe to inject an EM signal at four potential points, including the sensing array, a signal processing module (incorporating amplifiers, an ADC, and an MCU), a communication channel, and a communication module, as shown in \cref{fig-pre-inj}. To isolate each target, the other three points were shielded with aluminum foil. The shielding ensured that only the most sensitive point would be affected, preventing simultaneous coupling. In addition,  a copper mesh was used to shield the oscilloscope probe, thereby minimizing direct coupling of EMI signals into the probe. We performed sweep experiments over the range of 700 MHz to 1500 MHz at four points in turn and monitored the TXD of the universal asynchronous receiver/transmitter (UART) with an oscilloscope probe and the output signal using the host computer software in real-time to detect interference.
The results showed that the signal was successfully injected at two components:
\begin{itemize}
\item \textbf{Sensing array} (Point 1): At 701~MHz, injection at the sensing array caused a stable positive interference in the sensor measurement, indicating that the interference was caused by direct coupling with the sensor's electrodes. In addition, the sensor output deviation can be positive or negative depending on the EMI frequency, as shown in~\cref{fig-pre-experi}.

\item \textbf{Communication channel} (Point 3): At 1380~MHz, injection at the communication line resulted in data loss and signal amplitude fluctuations, causing a sensor DoS.
\end{itemize}

\begin{figure}[t]
    \centering
    \includegraphics[width=1.0\linewidth]{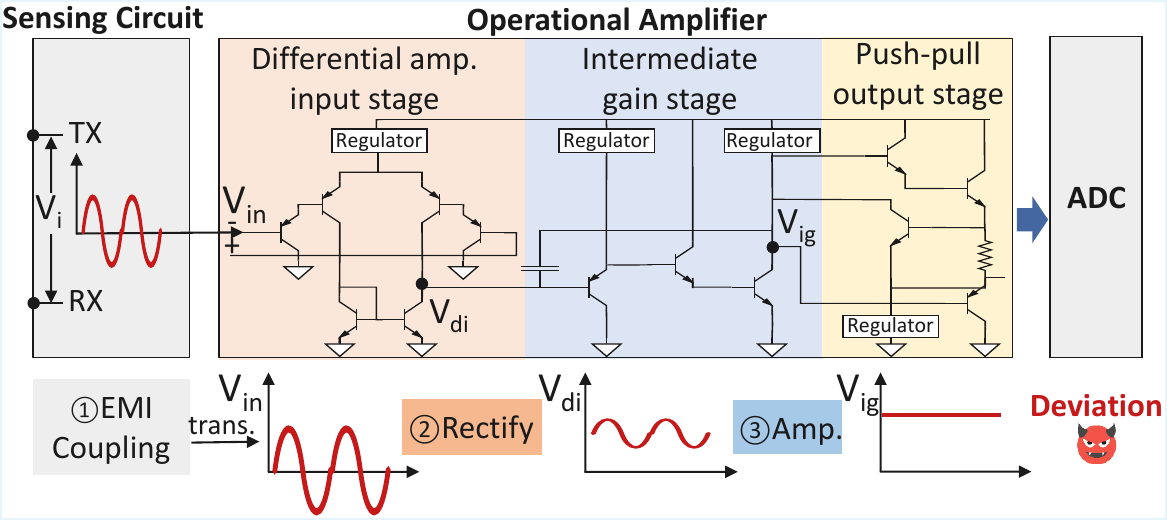}
    \caption{Principle of the injected EMI impact on op-amp. The EMI signal is coupled into the sensing circuits, then rectified and amplified by the op-amp, resulting in a DC offset.
    }
    \label{fig-pre-lm358}
\end{figure}

\subsection{Interference-induced Deviation Analysis}
This section explains and validates how high-frequency EMI signals induce stable deviations in sensor measurements.
Given the sensing array's susceptibility to EMI, the injected EMI signals are subsequently processed by an op-amp. As the primary signal conditioning stage before digitization, the op-amp is susceptible to the nonlinear distortion of high-frequency signals.
An op-amp typically comprises a differential input stage, an intermediate gain stage, and a push-pull output stage. When an EMI signal couples into the circuit as a high-frequency electrical signal, it impacts the op-amp's output through two main mechanisms: nonlinear rectification and bandwidth-limited amplification, as shown in~\cref{fig-pre-lm358}.

\subsubsection{Nonlinear Rectification}
The PN junctions within the bipolar junction transistors of the differential input stage drive the rectification process. These junctions efficiently convert high-frequency AC input signals into a DC component at the op-amp output due to their nonlinear current-voltage behavior~\cite{analog2009rfi}.

We model this rectification by applying a high-frequency input signal $v(t) = V_x \cos(2\pi f t)$ to the base-emitter junction of the operational amplifier. This input creates a voltage perturbation $\Delta V_{BE}$. The Shockley diode equation defines the static relationship between the collector current $I_c$ and the base emitter voltage $V_{BE}$ as $I_c= I_s(e^{V_{BE}/{V_T}}-1)$.  We analyze the impact of this perturbation using a Taylor expansion of $I_C$:
\begin{equation}
I_{C} \approx I_{C 0}+\frac{d I_{C}}{d V_{B E}} \Delta V_{B E}+\frac{1}{2} \frac{d^{2} I_{C}}{d V_{B E}^{2}}\left(\Delta V_{B E}\right)^{2}+\cdots
\end{equation}

The squared perturbation $(\Delta V_{BE})^2$ can be calculated as :
\begin{equation}
    (\Delta V_{BE})^2=V_x^2\cos^2(2\pi ft)=\frac{V_x^2}{2}+\frac{V_x^2}{2}\cos(4\pi ft)
\end{equation}
The second order term yields a DC component proportional to ${V_x^2}/{2}$, which can be formulated as:
\begin{equation} \label{eq_ic}
    \Delta I_{C(DC)}=\frac{1}{2}\frac{d^{2}I_{C}}{dV_{BE}^{2}} \frac{V_{x}^{2}}{2}=\frac{g_{m}^{\prime}V_{x}^{2}}{4}
\end{equation}
where the $g_{m}^{\prime}$, defined as $I_c / V^2_{T}$, quantifies the nonlinearity intensity and $V_T$ represents the thermal voltage constant. 

\begin{figure}[t]
    \centering
        \subfigure[BJT circuit. ]{
        \begin{minipage}[t]{0.33\linewidth}
            \centering
            \includegraphics[width=1\textwidth]{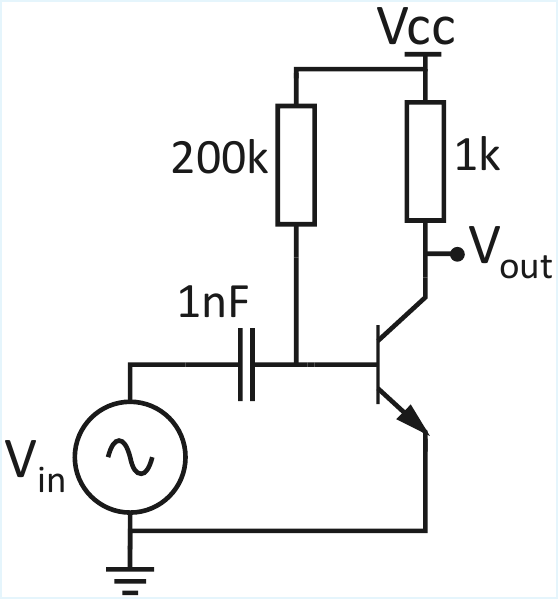}
        \end{minipage}
        \label{fig-pre-diode-sche}
    }
    \subfigure[Output signal at different input frequencies.]{
        \begin{minipage}[t]{0.57\linewidth}
            \centering
            \includegraphics[width=1\textwidth]{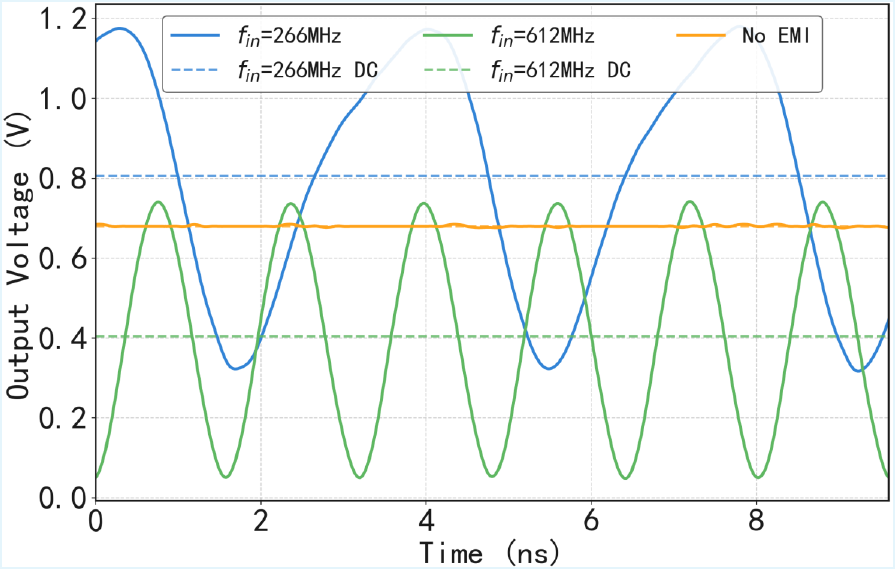}
        \end{minipage}
        \label{fig-pre-diode-wave}
    }

    \caption{The setup and outputs of the physical experiment for DPI in a BJT. The injected
high-frequency signal will generate a DC offset through nonlinear rectification.}
    \label{fig-pre-diode}
\end{figure}
In practical circuits, parasitic inductance and capacitance create a distributed impedance network with multiple signal paths. We analyze the parallel configuration as an example, where branches sum in the admittance domain. The ratio of net susceptance to net conductance across the $N$ parallel branches determines the frequency-dependent phase shift $\phi(f)$ :
\begin{equation}
\phi(f) = \arctan \left( \frac{ \sum_{k=1}^{N} \left( \frac{1}{2\pi f L_k} - 2\pi f C_k \right) }{ \sum_{k=1}^{N} \frac{1}{R_k} } \right)
\end{equation}
where $1/R_k$ represents the conductance. The numerator structure ($\frac{1}{\omega L} - \omega C$) reflects the phase inversion inherent in deriving impedance phase from total admittance since the impedance phase is the negative of the admittance phase.
The impedance phase $\phi(f)$ dictates the nonlinear response across two states. Near resonance the resistive regime promotes current pumping, which amplifies the average collector current and produces a negative output offset. Conversely, the capacitive behavior away from resonance induces base clamping where charge accumulation suppresses the base bias. This shifts the operating point and results in a positive output offset.

Rectification acts as a second-order phenomenon proportional to the signal power $V^2_x$. The effective nonlinearity arises from the balance between the resistive contribution $\cos^2\phi$ and the reactive contribution $\sin^2\phi$. Consequently the net equivalent transconductance $g_{m}^{\prime}(f)$ is defined as the difference between these components:
\begin{equation} \label{eq-g(f)}
g_{m}^{\prime}(f) = \frac{I_c}{V^2_{T}} \cos(2\phi(f))
\end{equation}
where the $\cos(2\phi(f))$ factor determines the offset polarity. At resonance, where $\phi$ is near zero, the positive value indicates that current pumping dominates. In the capacitive regime where $\phi$ approaches $\pm 90^\circ$, the term becomes negative because $\cos(\pm180^\circ)$ equals -1. This confirms that base clamping governs the response.

Finally, combining \cref{eq_ic} and \cref{eq-g(f)} yields the induced DC voltage offset $\Delta V_{DC}$:
\begin{equation} \label{eq-final-offset}
\begin{aligned}
    \Delta V_{DC} &= -\Delta I_{C(DC)} R_{r} = -\left( \frac{V_x}{V_{T}} \right)^{2} \frac{I_{C} R_{r}}{4} \cos(2\phi(f)) \\
    &= -\frac{I_{C} R_{r} R_{in}}{2 V_{T}^{2}} P_{i} \cos(2\phi(f))
\end{aligned}
\end{equation}
where $R_{in}$ and $R_r$ represent the equivalent input impedance and output load resistance, respectively. Rectification occurs at the differential amplification stage and produces a signal containing a DC component. The intermediate gain stage then processes this signal.

In summary, \textbf{nonlinear rectification of the high-frequency signal induces a DC offset, and the frequency determines the polarity of the resulting DC component.}

\textbf{Experimental validation.}
To verify the nonlinear rectification effect, we conducted direct power injection (DPI) experiments on the BJT-based circuit (shown in~\cref{fig-pre-diode}) by performing a frequency sweep.  The results reveal a +0.13 V DC offset superimposed on the 266~MHz high-frequency signal and a -0.27 V DC offset under the 612~MHz signal. These findings confirm that the BJT nonlinearity effectively rectifies the EMI component and validates the theoretical analysis of DC offset generation. To ensure measurement accuracy, we filtered extraneous AC components caused by EM coupling to the probe to isolate the true output waveform and ensure measurement accuracy.

\begin{figure}[t]
    \centering
        \subfigure[The experiment setup.]{
        \begin{minipage}[t]{1.0\linewidth}
            \centering
            \includegraphics[width=1\textwidth]{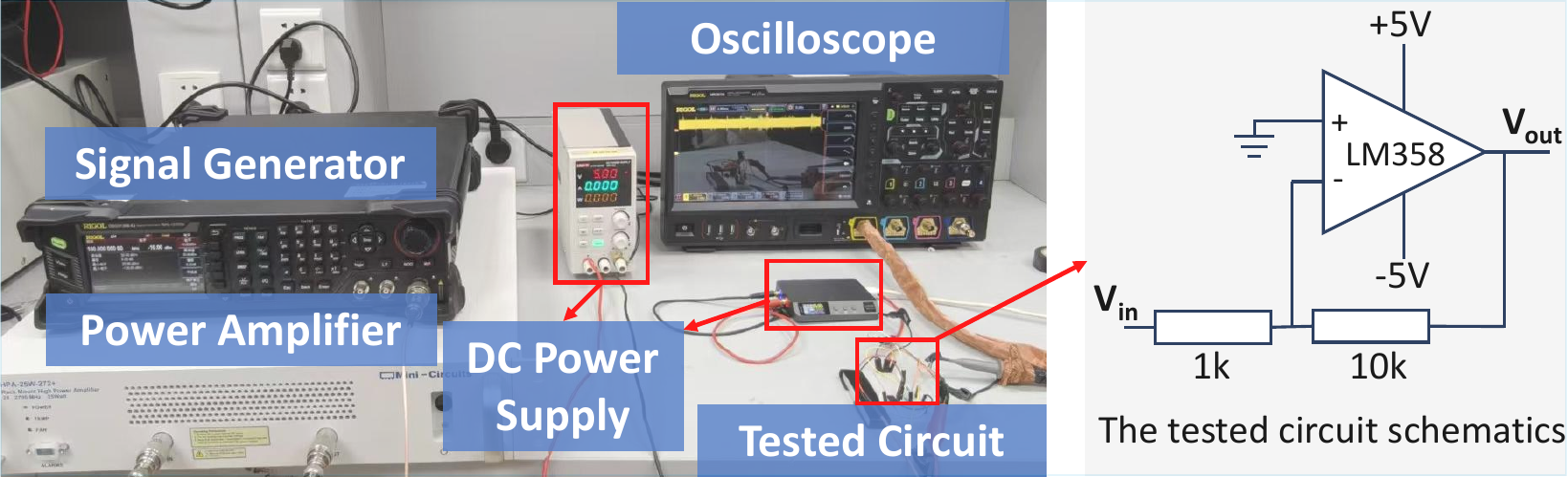}
        \end{minipage}
        \label{fig-pre-lm358-test}
    }
    \subfigure[The output signal waves under different input frequency.]{
        \begin{minipage}[t]{1.0\linewidth}
            \centering
            \includegraphics[width=1\textwidth]{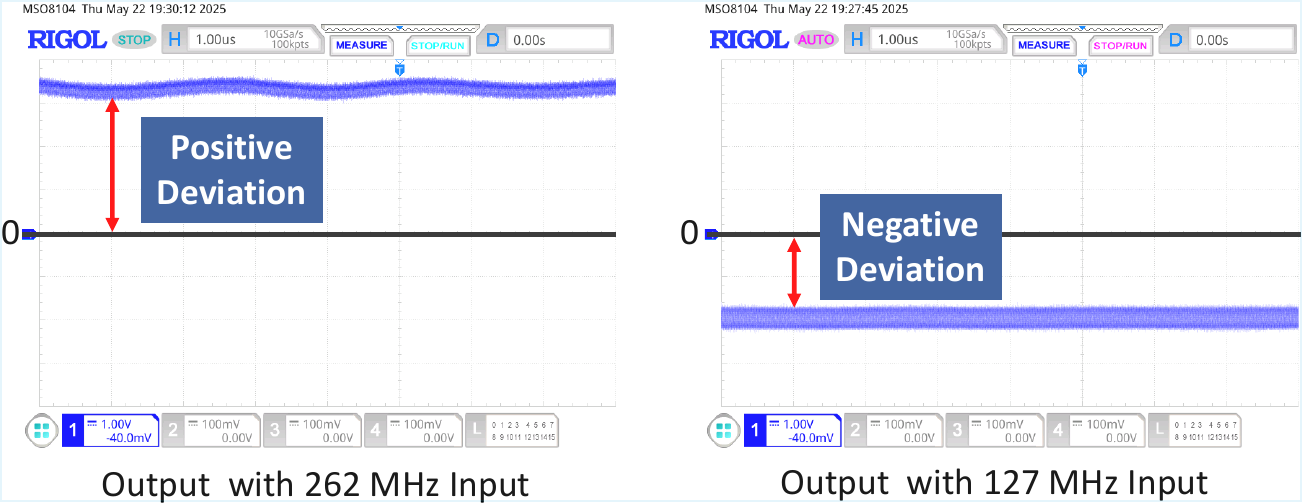}
        \end{minipage}
        \label{fig-pre-lm358-wave}
    }
    \caption{Real-world test using DPI on an op-amp. }
    \label{fig-pre-lm358-all}
\end{figure}

\begin{figure*}[t]
    \centering
    \includegraphics[width=0.98\linewidth]{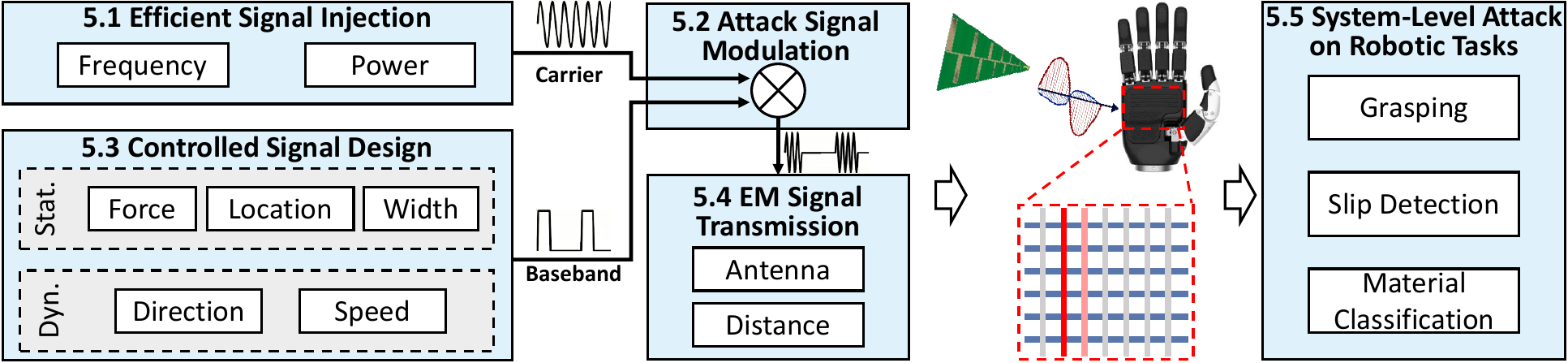}
    \caption{
    The workflow of the manipulation attack. The process begins with efficient signal injection, where the attacker optimizes carrier parameters (frequency and power) to maximize EM intensity. Simultaneously, controlled signal design is achieved by crafting baseband signals to precisely manipulate static (stat.) effects (force, location, and width) and dynamic (dyn.) effects. These components are combined via attack signal modulation and propagated through EM transmission. Ultimately, the EM signal couples into the target sensors, leveraging these mechanisms to execute attacks across diverse scenarios.}
    \label{fig-design-overview}
\end{figure*}
\subsubsection{Bandwidth-limited Amplification}
The amplification effect occurs in the intermediate gain stage. 
The transistors in the op-amp introduce parasitic capacitances.
We model the output impedance as a resistance $R_o$ in parallel with a total output capacitance $C_o$, which represents the combined parasitic effects at the output node. 
The gain of the intermediate stage can be formulated as:
\begin{equation}
    A_{v}(s)=\frac{V_{out}(s)}{V_{in}(s)}= \frac{-(g_{m}V_{in}(s)) \cdot R_{o} \parallel \frac{1}{sC_{o}}}{V_{in}(s)} = -\frac{g_{m} R_{o}}{1+s R_{o} C_{o}}
\end{equation}
where $s = j2\pi f$ and  $g_{m}V_{in}(s)$ denotes the controlled current source.

At low frequencies, $C_o$ exhibits a high reactance and can be approximated as an open circuit. As the frequency increases, the reactance of $C_o$ decreases, allowing more signal current to be shunted through the capacitive
path. The resulting decrease in effective output impedance lowers the magnitude of the voltage gain.

In summary, \textbf{the limited bandwidth of the op-amp attenuates high-frequency signals while the DC component undergoes amplification and interferes with sensor measurement.}

\textbf{Experimental validation.}
To eliminate the uncertainty of EM coupling paths and efficiency, we still used DPI to analyze the impact of EMI on the internal components of an op-amp. We selected the LM358 model, which is prevalent in tactile sensors due to its low power consumption and dual-channel configuration. These features make it ideal for signal conditioning \cite{zhu2022soft, cai2021multifunctional, chen2024rigid}.

To validate the theoretical analysis presented previously, we constructed an inverting amplifier circuit as shown in \cref{fig-pre-lm358-test}. We applied a 2 V $V_{pp}$ sinusoidal input $V_{i1}$ and monitored the output signal $V_{out}$. 
We conducted a frequency sweep and observed that specific EMI frequencies induce a significant DC offset. For instance, we observed a positive deviation at 262 MHz and a negative deviation at 127 MHz. Conversely, the corresponding AC signal remained below the noise floor, as shown in \cref{fig-pre-lm358-wave}. 
This demonstrates that an attacker can manipulate the sensor output by tuning the interference frequency and inject false signals that remain indistinguishable from legitimate measurements.

\subsection{DoS Effect Analysis}
\subsubsection{Principle Analysis of DoS Effect} 
EMI attacks induce currents in the target's wiring~\cite{kaur2011electromagnetic}, leading to DoS via two primary mechanisms.
First, high-intensity interference can corrupt communication frames, forcing the host software into a deadlock state that requires a restart~\cite{cho2016error}. Second, EMI can trip hardware protection latches, disabling the sensor until a hardware reboot~\cite{ti2024protect}.

\subsubsection{Experiment Validation of DoS Effect} \label{pre-DoS}

To validate our analysis, we tested a tactile sensor module~\cite{legact} (shown in~\cref{fig-pre-inj}) and a dexterous hand~\cite{hand}.
Frequency sweeps revealed that a 1380 MHz signal caused persistent communication loss. Waveform analysis of the TXD line showed severe data corruption during the attack, while communication fully recovered after a software reboot. This confirms corrupted protocol frames caused host parsing failures, which triggered the DoS effect rather than physical hardware damage.
In contrast, the commercial hand~\cite{hand} exhibited a hardware-level failure at a 1443 MHz signal. Attack onset caused sensor measurements to saturate at $65535\text{ g}$ momentarily before dropping to zero. This unresponsive state persisted after the attack stopped and restarting the host software failed to resolve the issue. Restoring functionality required a hardware reboot, indicating the interference triggered a persistent hardware latch rather than a software fault.

%% file: sections/design.tex
\section{\xxx Design} \label{Design}

Building on the established EMI coupling and interference principles, we present the design of \xxx, an attack framework for realizing contactless manipulation of tactile measurement. 
As illustrated in \cref{fig-design-overview}, \xxx jointly designs the carrier and baseband attack signals to achieve controllable interference over tactile sensor outputs, enabling regulation of perceived force, steering of perturbations to targeted locations, adjustment of the affected width, and synthesis of dynamic patterns. 
We further demonstrate how these controllable mechanisms can be composed to attack three representative dexterous hand applications, i.e., grasping, slip detection, and material classification.

\subsection{Efficient Signal Injection}

We first design a carrier signal that enables EMI signals to couple into the victim tactile sensor with high efficiency. In \xxx, we use a sinusoidal waveform as the carrier and select its key parameters, i.e., carrier frequency and transmission power, to maximize coupling strength and the resulting sensing perturbation.

\begin{figure}[t]
    \centering
    \includegraphics[width=1\linewidth]{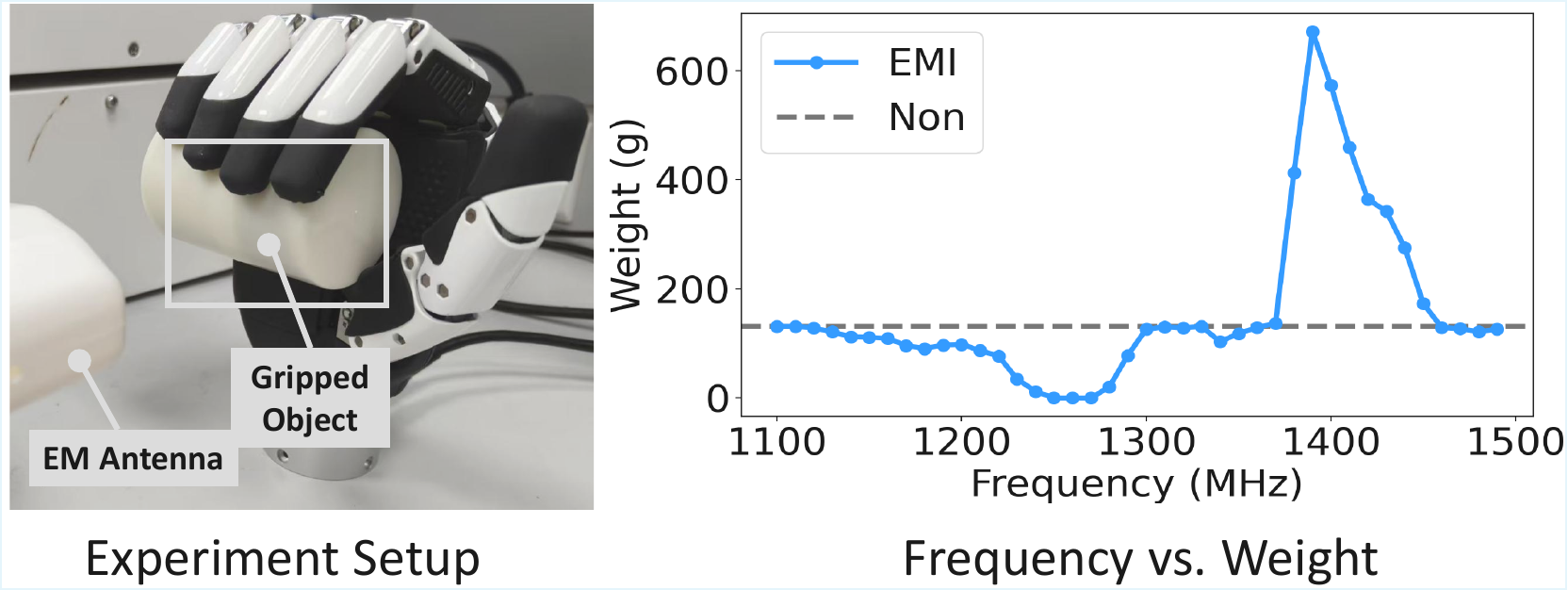}
    \caption{The impact of EM signal frequency on weight measurements.}
    \label{fig-design-fre}
    \vspace{-5pt}
\end{figure}

\subsubsection{Signal Frequency}
The coupling efficiency of an EM signal depends on how closely its frequency matches the electrical characteristics of the target conductor. In EMI attacks, the internal wiring of tactile sensors can unintentionally function as receiving antennas.
For a wire of length $l$, strong coupling typically occurs when the signal frequency satisfies ${c}/ 50l \leq f \leq {c}/{2l}$, where $c$ denotes the speed of light.
The exact resonant frequency further depends on conductor geometry, material properties, and surrounding circuitry. Although these factors are not directly observable, an effective operating frequency can be identified through a frequency sweep while monitoring the sensor response.

We validate this relationship using a tactile sensor on a dexterous hand~\cite{hand}.
With the sensor gripping an object and registering a baseline average of 200~g across all units (\cref{fig-design-fre}), we sweep the carrier frequency from 1.10~GHz to 1.50~GHz in 10~MHz increments and record the average reading of all sensing units. As shown in \cref{fig-design-fre}, coupling efficiency varies sharply with frequency. 1400~MHz produces the strongest positive interference, whereas the 1.25–1.27~GHz range yields the largest negative interference, consistent with theoretical expectations.

\begin{figure}[t]
    \centering
    \includegraphics[width=1\linewidth]{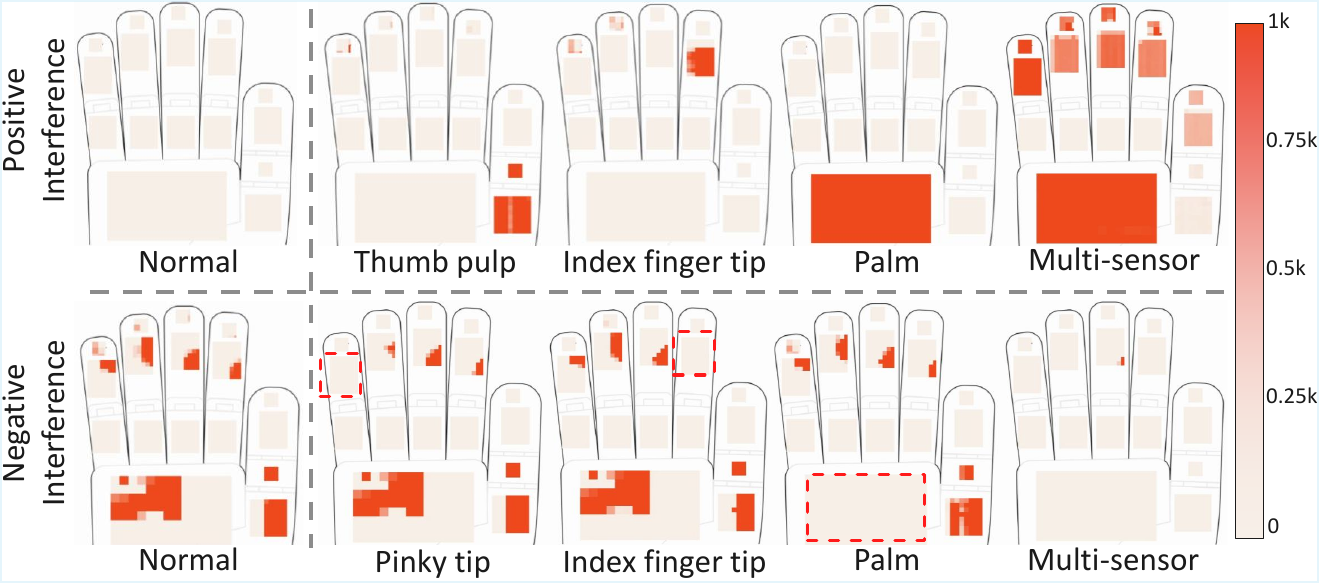}
    \caption{Target tactile sensor selection on dexterous hand via EMI signal frequency. Positive interference tests utilized an unloaded hand, whereas negative interference tests involved grasping an object.}
    \label{fig-design-fre-multi}
\end{figure}

Modern robotic systems integrate multiple tactile sensors whose resonant frequencies differ due to variations in sensor type, placement, and array geometry. This diversity enables selective targeting. By fine-tuning the injection frequency, an attacker can preferentially interfere with a chosen sensor while exerting minimal impact on adjacent ones. Furthermore, signals composed of multiple frequencies enable simultaneous interference with multiple sensors as demonstrated in \cref{fig-design-fre-multi}.

\begin{figure}[t]
    \centering
        \subfigure[Power vs. weight.]{
        \begin{minipage}[t]{0.438\linewidth}
            \centering
            \includegraphics[width=1\textwidth]{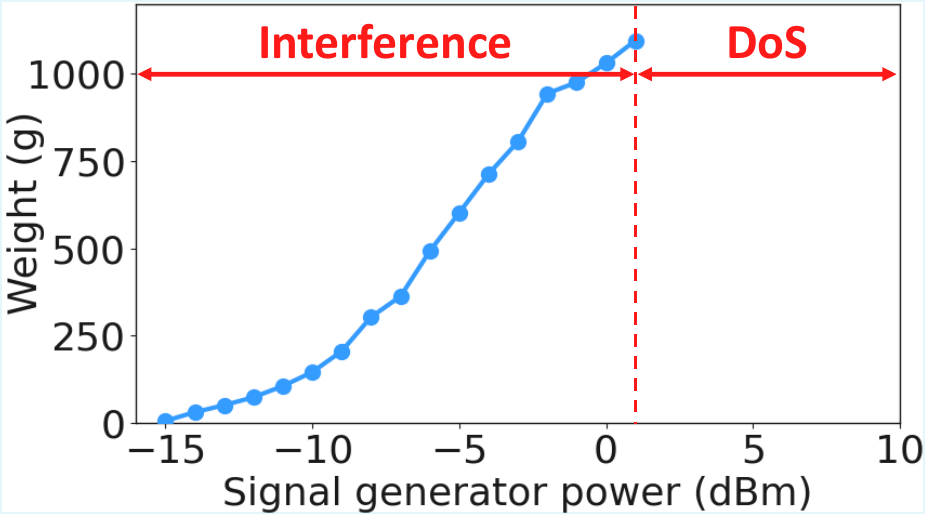}
        \end{minipage}
        \label{fig-design-amp1}
        
    }
    \subfigure[Power vs. intensity.]{
        \begin{minipage}[t]{0.462\linewidth}
            \centering
            \includegraphics[width=1\textwidth]{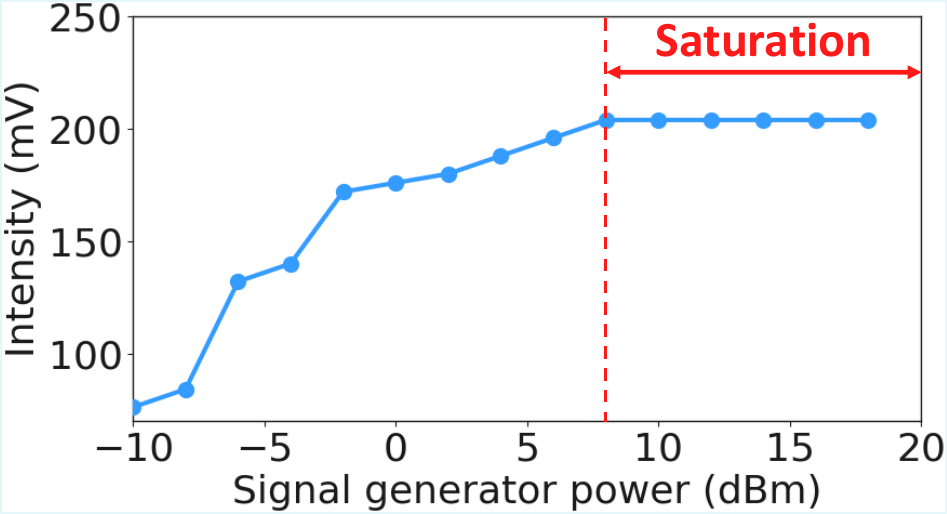}
        \end{minipage}
        \label{fig-design-amp2}
        
    }
    \label{fig-design-amp}
    \caption{The impact of signal power on injected weight for sensors and voltage magnitude for the probe. }
\end{figure}

\subsubsection{Signal Power}

Transmission power directly influences injection effectiveness. In general, higher transmission power yields stronger interference, while the received electromagnetic energy decreases approximately with the inverse square of the
transmission distance~\cite{weik2000computer}. According to~\cref{eq-final-offset}, the induced force variation is approximately proportional to the injected signal power.

We sweep the transmission power of a 1443~MHz carrier from −15~dBm to 10~dBm and measure the corresponding weight variation at 1~dBm intervals. As shown in \cref{fig-design-amp1}, the weight deviation increases approximately proportionally with power. When the output exceeds 2~dBm, the tactile sensor enters a DoS state.

Practical constraints limit further gains from increasing transmission power. Measurements using a near-field probe (\cref{fig-design-amp2}) reveal that the power amplifier saturates at approximately 10~dBm input, beyond which no additional output increase is observed. This highlights the importance of selecting carrier power within the effective operating range of the attack hardware.

\subsection{Attack Signal Modulation}
While a single-frequency carrier can disrupt sensor output, precise manipulation requires embedding a structured baseband signal. Common modulation schemes include frequency modulation (FM), phase modulation (PM), and amplitude modulation (AM).

FM perturbs the target sensor by varying the carrier frequency. However, sensor responses exhibit strong frequency dependence and unpredictability, and effective FM requires detailed knowledge of the sensor’s frequency characteristics, leading to high attack complexity. PM modulates the carrier phase but produces highly device-dependent and often nonlinear sensor responses, preventing a reliable mapping between injected signals and sensing effects. Therefore, neither FM nor PM is suitable for fine-grained and controllable manipulation in our setting.

In contrast, AM provides predictable and stable controllability. The induced sensor offset is approximately proportional to the amplitude of the injected EMI signal, enabling fine-grained control over sensor behavior.
Accordingly, we adopt AM as the modulation scheme in GhostTac. The transmitted AM signal is given by
\begin{equation}
s_{AM}(t)=(A_c + m(t))cos\left(2\pi f_ct \right)
	\label{eq_modulation}
\end{equation}
where $m(t)$ is the baseband, $A_c$ and $f_c$ denote the carrier amplitude and frequency.
In practice, $f_c$ targets the victim sensor resonant frequency for efficient coupling.  $A_c$ maximizes injected intensity and the baseband regulates the sensing perturbation.

\subsection{Controlled Signal Design} \label{sec-design-effects}

We present how \xxx generates controllable sensing perturbations by parameterizing the baseband signal ($m(t)$).
The attack configures four key parameters, i.e., baseband frequency $f_b$, amplitude $A_b$, temporal offset $\Delta t$, and duration $\Delta T$.
Together, we can realize two categories of attack, 1) \textbf{static manipulation}, which controls force, location, and affected width, and 2) \textbf{dynamic manipulation}, which synthesizes directional slipping effects.

\begin{figure}[t]
    \centering
    \includegraphics[width=1.0\linewidth]{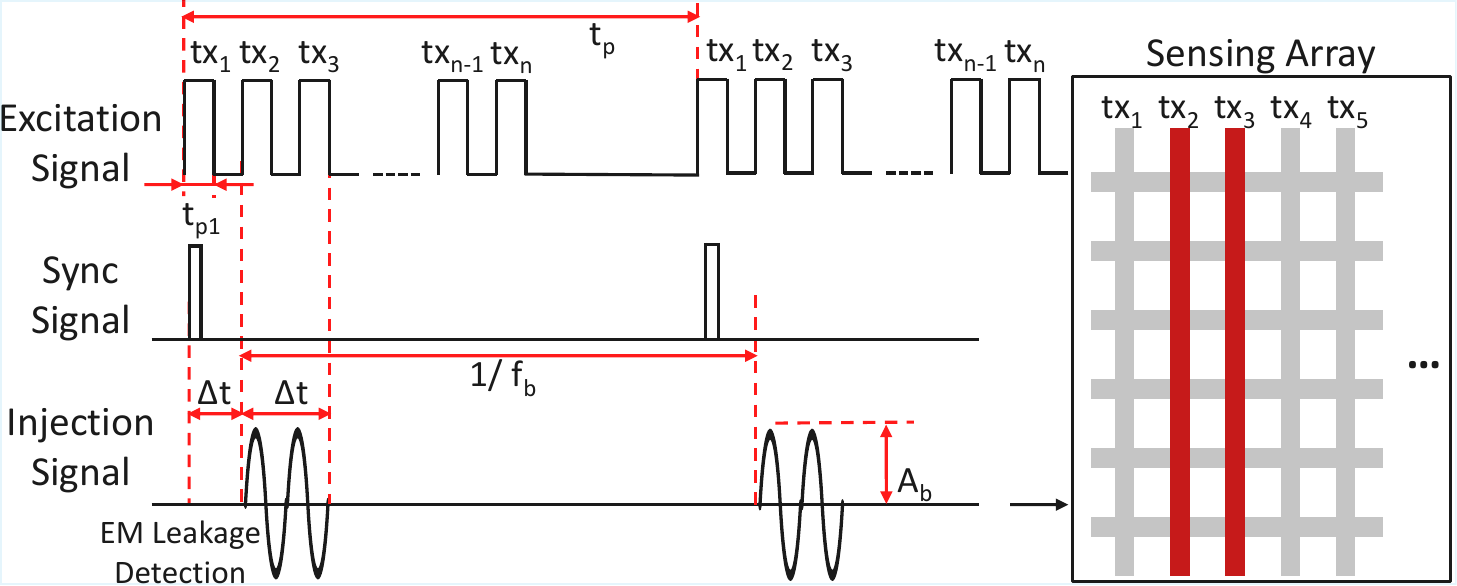}
    \caption{The signal injection strategy and manipulation of sensing units. Referring to the excitation signal and the timing, an attacker can interfere with certain TXs. }
    \label{fig-design-syn}
\end{figure}

\subsubsection{Force}
\begin{figure}[t]
    \centering
    \includegraphics[width=1.0\linewidth]{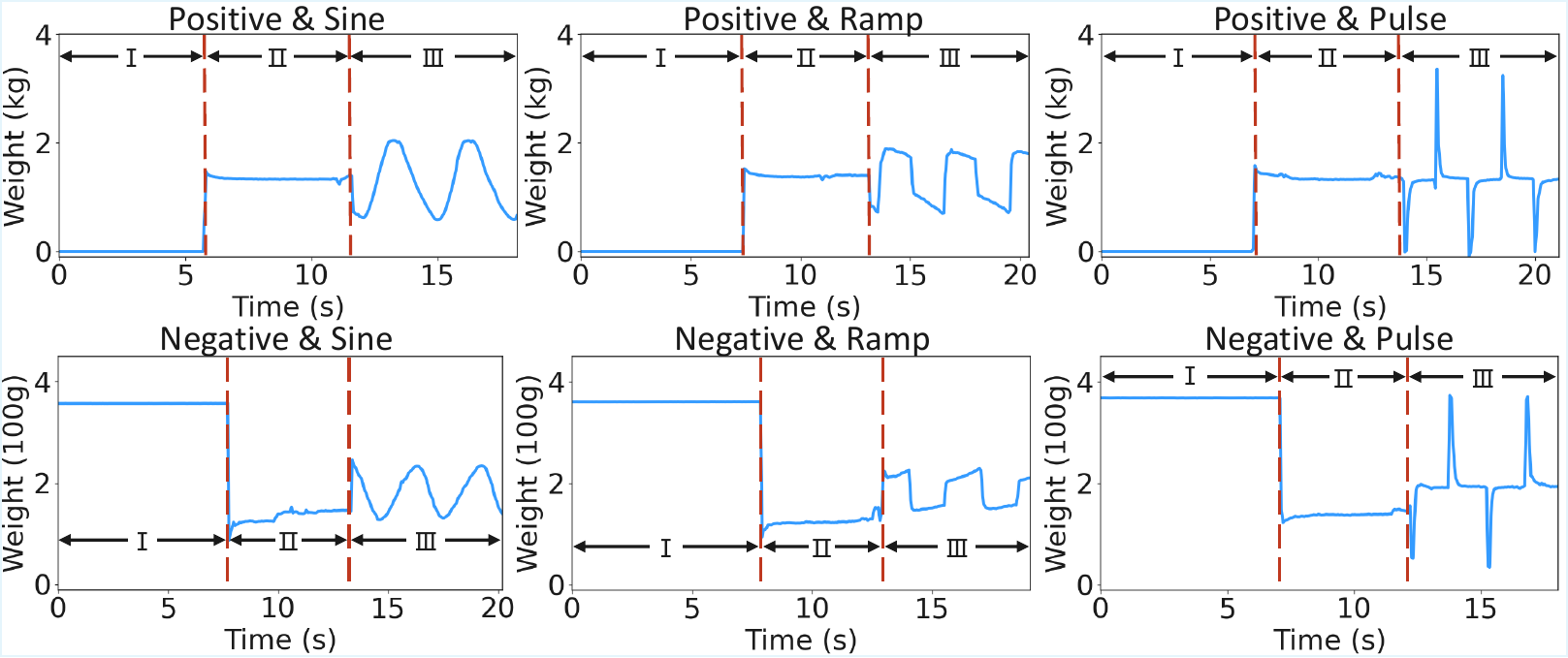}
    \caption{The sensor measurements under three situations. \textrm{I}: Without EMI; \textrm{II}: Single-frequency EMI; \textrm{III}: AM-based EMI.}
    \label{fig-des-modulate}
\end{figure}

Force manipulation is achieved by regulating the amplitude of the injected baseband signal. Since tactile sensors map signal magnitude to applied pressure, adjusting the baseband amplitude $A_b$ directly controls the perceived force.
For static manipulation, the baseband frequency $f_b$ is aligned with the sensor’s excitation scanning frequency to ensure stable force injection across successive sensing cycles.
We validate this capability by injecting baseband signals with sine, ramp, and pulse envelopes. As shown in \cref{fig-des-modulate}, the resulting sensor outputs closely track the injected waveforms, demonstrating that \xxx can synthesize diverse force profiles, including gradual compression and vibration.

\subsubsection{Location}
Location control enables the attacker to inject false measurements into specific sensing positions on the tactile array. This is achieved by synchronizing the interference signal with the sensor’s excitation process. By monitoring the electromagnetic leakage generated during excitation, the attacker can determine in real time which TX electrode is currently being processed.
The attacker then exploits the latency $\Delta t$ between the captured synchronization signal and the target excitation event to align the injection with a specific TX. 
The required delay to target the $i$-th electrode is given by:
\begin{equation}
    \Delta t = (i-1)T_{p1} + T_{de}
\end{equation}
where $T_{p1}$ represents the duration of a single TX activation and $T_{de}$ accounts for transmission latency. 
By varying $\Delta t$, the attacker can shift the interference region across the tactile array. 
In practice, attackers may employ a hardware-identical device to recover $\Delta t$, as shown in~\cref{fig-design-syn}.
If reliable EM leakage is unavailable, the attacker may alternatively set $\Delta t$ relative to the baseband period, thereby injecting perturbations at unpredicted but fixed locations on the tactile array.

\subsubsection{Width}
\begin{figure}[t]
   \centering
   \begin{minipage}[b]{.455\linewidth}
      \includegraphics[width=1.\linewidth]{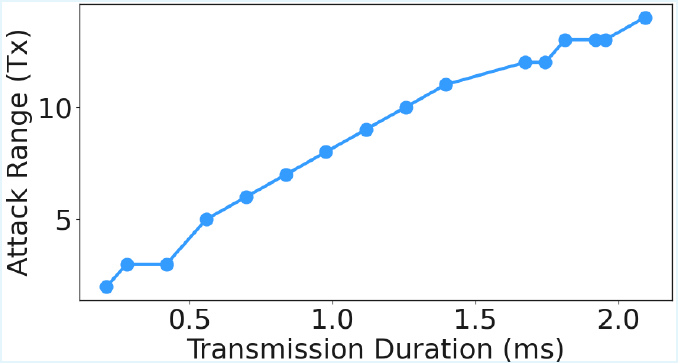}
      \caption{The interference width as $\Delta T$ increases.}
      \label{fig-design-duty}
      
   \end{minipage}\qquad
   \begin{minipage}[b]{0.455\linewidth}
      \includegraphics[width=1.\linewidth]{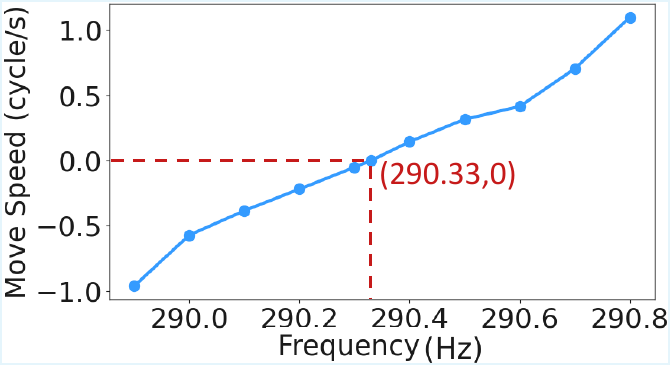}
      \caption{The movement speed as $f_b$ changes.}
      \label{fig-design-speed}
   \end{minipage}
\end{figure}

Width manipulation controls the spatial extent of the injected interference and does not require synchronization with the excitation process.
After fixing the injection start position, the number of TX electrodes covered by the interference region is determined by the signal duration $\Delta T$.
Increasing $\Delta T$ activates a larger contiguous region of sensing units.
We sweep the duty cycle of a modulated square-wave injection from 0.06 to 0.60 and record the resulting interference range. The results in \cref{fig-design-duty} show that the attacked width increases monotonically with $\Delta T$.

\subsubsection{Dynamic Manipulation}
Dynamic manipulation synthesizes moving interference patterns that simulate slipping. 
Unlike static injection, this strategy intentionally introduces a small mismatch between the baseband period $T_b$ and the sensor scanning period $T_p$.
This mismatch causes the injection phase to drift relative to the scanning window across successive frames, producing apparent spatial motion of the interference region.
The magnitude of the mismatch determines the speed of motion, while its sign determines the direction. In our experiments, we vary the modulated square-wave frequency from 290.8 Hz to 289.9 Hz in 0.1 Hz steps and measure the resulting speed of the affected sensing units. As shown in \cref{fig-design-speed}, this approach enables controlled motion of simulated contact in either direction along the TX-controlled axis and at configurable speeds.

\subsection{EM Signal Transmission}
\begin{figure}[t]
    \centering
        \subfigure[EM signal emitters.]{
        \begin{minipage}[t]{0.4\linewidth}
            \centering
            \includegraphics[width=1\textwidth]{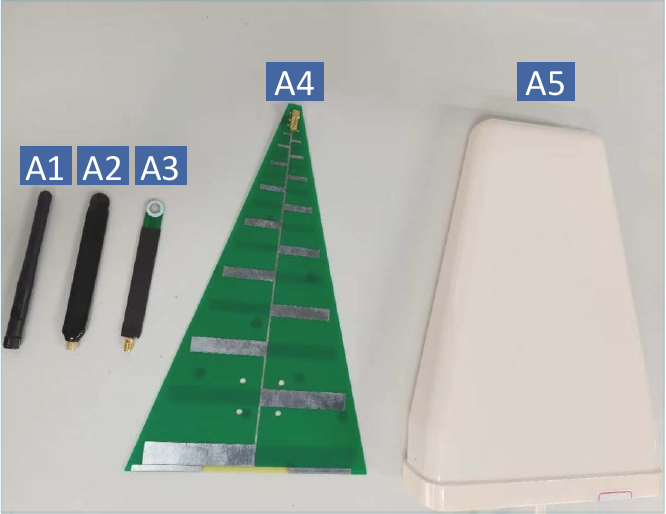}
        \end{minipage}
        \label{fig-design-attena1}
        
    }
    \subfigure[Comparison of five antennas.]{
        \begin{minipage}[t]{0.49\linewidth}
            \centering
            \includegraphics[width=1\textwidth]{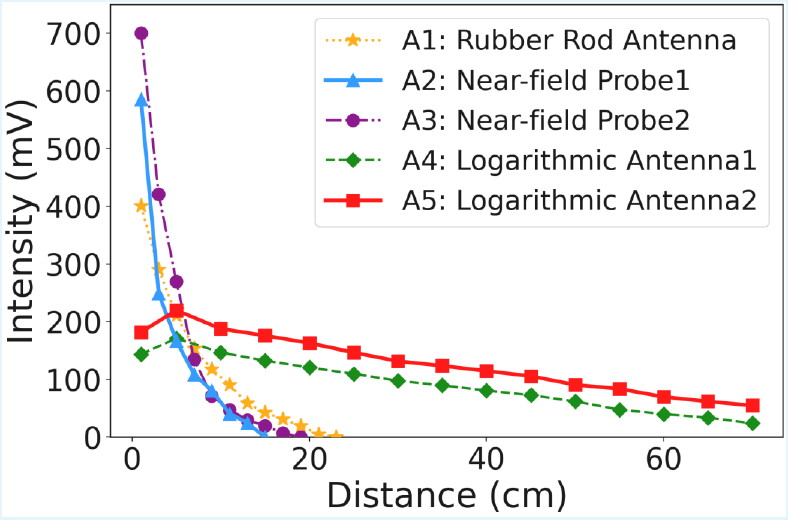}
        \end{minipage}
        \label{fig-design-attena2}
        
    }
    \label{fig-design-attena}
    \caption{The injected intensity of five EM signal antennas as the attack distance increases. }

\end{figure}

In \xxx, the attack signals are radiated through space via a transmission antenna. To achieve effective injection, the attacker seeks high antenna gain and proper impedance matching with the power amplifier to maximize delivered interference power and avoid hardware instability.
We evaluate multiple antenna types in~\cref{fig-design-attena1}, including rubber rod antennas, near-field probes, and log-periodic antennas, and observe distinct operational regimes. To compare their performance, we injected EM signals with each antenna and measured the received signal intensity with a separate electromagnetic probe at distances from 1 cm to 70 cm (\cref{fig-design-attena2}).

Near-field probes provide strong coupling at very short distances but suffer from rapid attenuation, making them suitable for close-range attacks. In contrast, log-periodic antennas offer superior directionality and maintain coupling strength over longer distances, making them more effective for non-contact injection at moderate ranges~\cite{gregson2007principles}. 
Interestingly, the log-periodic antenna exhibits stronger coupling as the distance increases from 1 cm to 5 cm. This is driven by near-field magnetic dynamics, where the field transitions from a disordered, surface-bound state to a spatially coherent structure that facilitates effective coupling~\cite{gregson2007principles}.

Based on these observations, GhostTac selects the log-periodic antenna when the attack distance exceeds 5~cm, and otherwise employs the near-field probe for close-proximity injection.

\subsection{System-Level Attack on Robotic Tasks}
We now demonstrate how fine-grained manipulation capabilities of  \xxx translate into practical attacks on representative dexterous hand-based tasks. By composing the primitive control dimensions, namely force, location, width, and dynamic timing, an attacker can systematically compromise key robotic functionalities that rely on tactile sensing.

\subsubsection{Grasping}
Tactile closed-loop grasping dynamically adjusts grip force based on real-time feedback~\cite{hogan2020tactile,romano2011human}. 
This adaptive process ensures stable handling of diverse objects~\cite{dean2017tomm} and prevents failures such as crushing caused by excessive compensatory force~\cite{kappassov2015tactile,luo2017robotic}. The correctness of tactile feedback is therefore fundamental to grasping stability and object safety.

\xxx directly targets this feedback loop. The attacker first identifies the carrier frequency $f_c$ that maximizes coupling efficiency using the frequency-sweep procedure.
On top of this carrier, the attacker modulates the baseband amplitude $A_b$ to precisely control the perceived contact force.
Through this parameterization, the attacker can steer the closed-loop controller toward unsafe operating states, such as inducing excessive grip force that damages fragile objects or weakening the grasp to cause object drop.

\subsubsection{Slip Detection}
Beyond passive closed-loop control, robots frequently employ active exploratory strategies for manipulation. A representative example is slip detection through motion-based probing, analogous to human hefting, in which the system deliberately induces inertial forces to detect incipient slippage. Such slip detection enables robots to perceive grasp instability and respond in real time~\cite{james2018slip}, enhancing autonomy and safety across diverse tasks~\cite{chen2018tactile}. Its failure can lead to object slip and task errors caused by insufficient grip~\cite{li2020review}.

Typical slip detection pipelines rely on monitoring temporal variations in force magnitude and spatial pressure distribution. \xxx exploits this mechanism by synthesizing dynamic interference patterns using baseband parameters.
After selecting the carrier $f_c$, the attacker configures the baseband frequency $f_b$ to be slightly offset from the sensor's scanning frequency, thereby creating a controlled temporal drift. This drift produces a moving interference pattern across the tactile array.
The attacker further sets the signal duration $\Delta T$ to match the object’s contact width and adjusts the frequency mismatch $\mid f_b-f_s\mid$ to control the apparent slip velocity. Higher mismatch produces faster simulated slip and triggers stronger compensatory responses. Using this construction, the attacker can either mask true slip events or induce false slip detection, leading to unstable or damaging grasp behavior.

\subsubsection{Material Classification}
Tactile material classification infers physical properties such as friction and compliance~\cite{jamali2010material}, enabling robots to adapt their grip strategy for safe and effective manipulation~\cite{li2020skin,sun2016object}. These pipelines integrate multiple tactile features, including force magnitude, contact location, and pressure distribution over time.

\xxx attacks this functionality by treating force, width, and location as tunable control variables under both static and dynamic interference conditions.
Concretely, the attacker modulates $A_b$ to shape force features, selects $\Delta t$ to target specific taxels on the array, and sets $\Delta T$ to control the spatial footprint of the injected perturbation.
These parameters are optimized offline in the digital domain to maximize classification error, and then physically instantiated using the corresponding EMI signals. This procedure yields consistent and robust misclassification under real-world conditions.
The effectiveness of these attacks is further examined in \cref{sec:case}, where we present case studies together with promising experimental results that illustrate their practical impact.

%% file: sections/Evaluation.tex
\section{Evaluation} \label{sec:case}

\subsection{Experiment Setup}

\begin{figure}[t]
    \centering
    \includegraphics[width=1.0\linewidth]{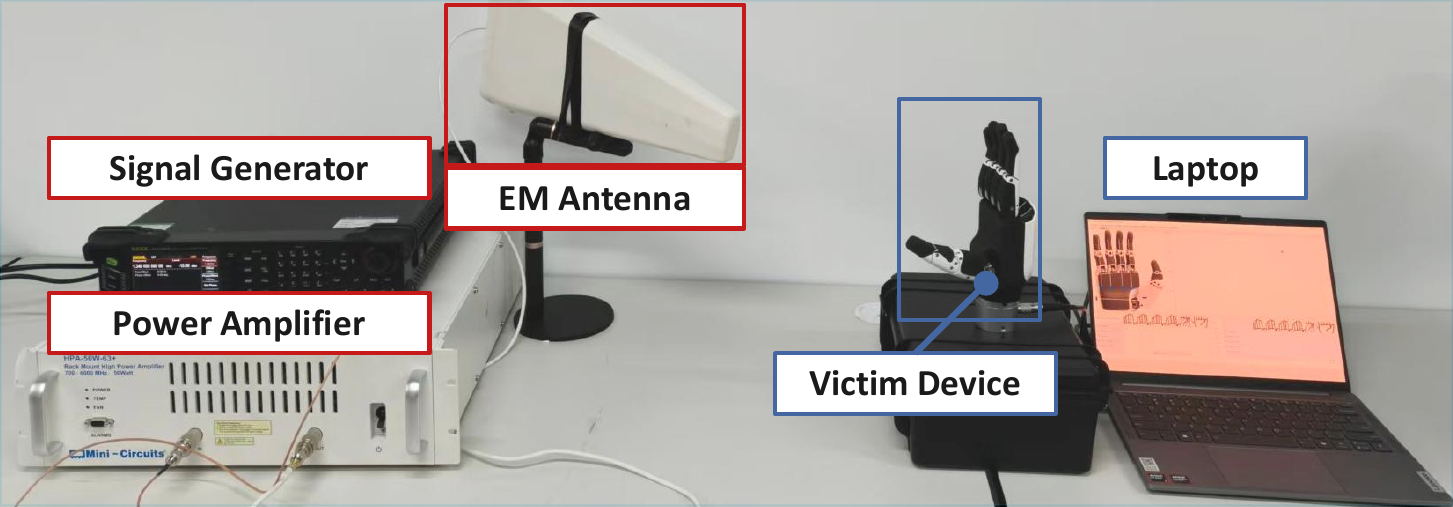}
    \caption{Physical implementation of the experimental setup for evaluating the performance of \xxx attack.}
    \label{fig-setup1}
\end{figure}

The \xxx attack system consists of a high-frequency signal generator~\cite{rigol2019} and a power amplifier~\cite{Mini-Circuits} with a maximum output of 50 W (47 dBm) to amplify high-frequency signals, which drive an EM emitter, a log-periodic antenna, or a near-field probe, as shown in~\cref{fig-setup1}.
Due to the weak power of the signal generator, the amplifier is used to boost the signal’s power before transmitting it into the EM emitter. The emitted EM signal will be injected into the victim device, causing sensor failures. 

\begin{table*}
    \centering

    \caption{Overall performance of \xxx on 15 tactile sensors.}
    \label{tab-perform}
    \renewcommand\arraystretch{1}
\resizebox{\linewidth}{!}{
\begin{tabular}{|ccccc|ccc|ccc|cccc|cc|}
\hline
\multicolumn{5}{|c|}{\multirow{2}{*}{\textbf{Sensor Information}}} & \multicolumn{3}{c|}{\textbf{Positive Interference}} & \multicolumn{3}{c|}{\textbf{Negative Interference}} & \multicolumn{4}{c|}{\textbf{Manipulation Types}} & \multicolumn{2}{c|}{\textbf{DoS}} \\ \cline{6-17} 
\multicolumn{5}{|c|}{} & \multicolumn{1}{c|}{\multirow{2}{*}{\begin{tabular}[c]{@{}c@{}}\textbf{Fre.}\\ \textbf{/MHz}\end{tabular}}} & \multicolumn{2}{c|}{\textbf{Deviation/g}} & \multicolumn{1}{c|}{\multirow{2}{*}{\begin{tabular}[c]{@{}c@{}}\textbf{Fre.}\\ \textbf{/MHz}\end{tabular}}} & \multicolumn{2}{c|}{\textbf{Deviation/g}} & \multicolumn{3}{c|}{\textbf{Static}} & \textbf{Dynamic} & \multicolumn{1}{c|}{\multirow{2}{*}{\begin{tabular}[c]{@{}c@{}}\textbf{Fre.}\\ \textbf{/MHz}\end{tabular}}} & \multirow{2}{*}{\begin{tabular}[c]{@{}c@{}}\textbf{Success}\\ \textbf{Rate}\end{tabular}} \\ \cline{1-5} \cline{7-8} \cline{10-15}
\multicolumn{1}{|c|}{\textbf{\#}} & \multicolumn{1}{c|}{\textbf{Manuf.}} & \multicolumn{1}{c|}{\textbf{Model}} & \multicolumn{1}{c|}{\textbf{Type}} & \textbf{Array} & \multicolumn{1}{c|}{} & \multicolumn{1}{c|}{\textbf{Max.}} & \textbf{Avg.} & \multicolumn{1}{c|}{} & \multicolumn{1}{c|}{\textbf{Max.}} & \textbf{Avg.} & \multicolumn{1}{c|}{\textbf{Force}} & \multicolumn{1}{c|}{\textbf{Location}} & \multicolumn{1}{c|}{\textbf{Width}} & \textbf{Slip} & \multicolumn{1}{c|}{} & \\ \hline\hline
\multicolumn{1}{|c|}{1} & \multicolumn{1}{c|}{\multirow{3}{*}{LEGACT}} & \multicolumn{1}{c|}{RPPS-36} & \multicolumn{1}{c|}{R} & 6*6 & \multicolumn{1}{c|}{1156} & \multicolumn{1}{c|}{+4870} & +4812 & \multicolumn{1}{c|}{1417} & \multicolumn{1}{c|}{-1226} & -374 & \multicolumn{1}{c|}{$\sqrt{}$} & \multicolumn{1}{c|}{$\sqrt{}$} & \multicolumn{1}{c|}{$\sqrt{}$} & $\sqrt{}$ & \multicolumn{1}{c|}{1380} & 10/10 \\ \cline{1-1} \cline{3-17} 
\multicolumn{1}{|c|}{2} & \multicolumn{1}{c|}{} & \multicolumn{1}{c|}{RPPS-32} & \multicolumn{1}{c|}{R} & 4*8 & \multicolumn{1}{c|}{705} & \multicolumn{1}{c|}{+982} & +303 & \multicolumn{1}{c|}{1280} & \multicolumn{1}{c|}{-192} & -60 & \multicolumn{1}{c|}{$\sqrt{}$} & \multicolumn{1}{c|}{$\sqrt{}$} & \multicolumn{1}{c|}{$\sqrt{}$} & $\sqrt{}$ & \multicolumn{1}{c|}{1325} & 10/10 \\ \cline{1-1} \cline{3-17} 
\multicolumn{1}{|c|}{3} & \multicolumn{1}{c|}{} & \multicolumn{1}{c|}{RP-S40} & \multicolumn{1}{c|}{R} & 1*1 & \multicolumn{1}{c|}{1130} & \multicolumn{1}{c|}{+802} & +802 & \multicolumn{1}{c|}{1090} & \multicolumn{1}{c|}{-490} & -490 & \multicolumn{1}{c|}{$\sqrt{}$} & \multicolumn{3}{c|}{N/A} & \multicolumn{1}{c|}{1310} & 9/10 \\ \hline
\multicolumn{1}{|c|}{4} & \multicolumn{1}{c|}{RouXi} & \multicolumn{1}{c|}{M0505S} & \multicolumn{1}{c|}{R} & 5*5 & \multicolumn{1}{c|}{701} & \multicolumn{1}{c|}{+120} & +56 & \multicolumn{1}{c|}{1190} & \multicolumn{1}{c|}{-520} & -437 & \multicolumn{1}{c|}{$\sqrt{}$} & \multicolumn{1}{c|}{$\sqrt{}$} & \multicolumn{1}{c|}{$\sqrt{}$} & $\sqrt{}$ & \multicolumn{2}{c|}{$\times$} \\ \hline
\multicolumn{1}{|c|}{5} & \multicolumn{1}{c|}{\multirow{3}{*}{Crownto}} & \multicolumn{1}{c|}{Sole} & \multicolumn{1}{c|}{R} & 4*6 & \multicolumn{1}{c|}{1025} & \multicolumn{1}{c|}{+131k} & +127k & \multicolumn{1}{c|}{1400} & \multicolumn{1}{c|}{-15k} & -9k & \multicolumn{1}{c|}{$\sqrt{}$} & \multicolumn{1}{c|}{$\sqrt{}$} & \multicolumn{1}{c|}{$\sqrt{}$} & $\sqrt{}$ & \multicolumn{1}{c|}{1185} & 10/10 \\ \cline{1-1} \cline{3-17} 
\multicolumn{1}{|c|}{6} & \multicolumn{1}{c|}{} & \multicolumn{1}{c|}{04-CK} & \multicolumn{1}{c|}{R} & 4*4 & \multicolumn{1}{c|}{800} & \multicolumn{1}{c|}{+81} & +49 & \multicolumn{1}{c|}{1062} & \multicolumn{1}{c|}{-21} & -11 & \multicolumn{1}{c|}{$\sqrt{}$} & \multicolumn{1}{c|}{$\sqrt{}$} & \multicolumn{1}{c|}{$\sqrt{}$} & $\sqrt{}$ & \multicolumn{1}{c|}{1350} & 10/10 \\ \cline{1-1} \cline{3-17} 
\multicolumn{1}{|c|}{7} & \multicolumn{1}{c|}{} & \multicolumn{1}{c|}{M1616S} & \multicolumn{1}{c|}{R} & 16*16 & \multicolumn{1}{c|}{1475} & \multicolumn{1}{c|}{+6} & +5 & \multicolumn{1}{c|}{1060} & \multicolumn{1}{c|}{-345} & -342 & \multicolumn{1}{c|}{$\sqrt{}$} & \multicolumn{1}{c|}{$\sqrt{}$} & \multicolumn{1}{c|}{$\sqrt{}$} & $\sqrt{}$ & \multicolumn{1}{c|}{1485} & 10/10 \\ \hline
\multicolumn{1}{|c|}{8} & \multicolumn{1}{c|}{Tekscan} & \multicolumn{1}{c|}{A401} & \multicolumn{1}{c|}{R} & 1*1 & \multicolumn{1}{c|}{841} & \multicolumn{1}{c|}{+811} & +811 & \multicolumn{1}{c|}{958} & \multicolumn{1}{c|}{-277} & -277 & \multicolumn{1}{c|}{$\sqrt{}$} & \multicolumn{3}{c|}{N/A} & \multicolumn{1}{c|}{1472} & 10/10 \\ \hline
\multicolumn{1}{|c|}{9} & \multicolumn{1}{c|}{HangKai} & \multicolumn{1}{c|}{Fingertip} & \multicolumn{1}{c|}{C} & 1*3 & \multicolumn{1}{c|}{1112} & \multicolumn{1}{c|}{+228} & +223 & \multicolumn{1}{c|}{1400} & \multicolumn{1}{c|}{-133} & -132 & \multicolumn{1}{c|}{$\sqrt{}$} & \multicolumn{3}{c|}{N/A} & \multicolumn{1}{c|}{1382} & 7/10 \\ \hline
\multicolumn{1}{|c|}{10} & \multicolumn{1}{c|}{Isense} & \multicolumn{1}{c|}{UM1208} & \multicolumn{1}{c|}{R} & 12*8 & \multicolumn{1}{c|}{1372} & \multicolumn{1}{c|}{+135} & +130 & \multicolumn{1}{c|}{1419} & \multicolumn{1}{c|}{-5} & -3 & \multicolumn{1}{c|}{$\sqrt{}$} & \multicolumn{1}{c|}{$\sqrt{}$} & \multicolumn{1}{c|}{$\sqrt{}$} & $\sqrt{}$ & \multicolumn{1}{c|}{1480} & 6/10 \\ \hline
\multicolumn{1}{|c|}{11} & \multicolumn{1}{c|}{Linker} & \multicolumn{1}{c|}{L10 Tip} & \multicolumn{1}{c|}{R} & 6*12 & \multicolumn{1}{c|}{1571} & \multicolumn{1}{c|}{+520} & +483 & \multicolumn{1}{c|}{1184} & \multicolumn{1}{c|}{-189} & -99 & \multicolumn{1}{c|}{$\sqrt{}$} & \multicolumn{1}{c|}{$\sqrt{}$} & \multicolumn{1}{c|}{$\sqrt{}$} & $\sqrt{}$ & \multicolumn{2}{c|}{$\times$} \\ \hline
\multicolumn{1}{|c|}{12} & \multicolumn{1}{c|}{\multirow{4}{*}{\begin{tabular}[c]{@{}c@{}}Inspire\\ FTP\end{tabular}}} & \multicolumn{1}{c|}{End} & \multicolumn{1}{c|}{R} & 3*3 & \multicolumn{1}{c|}{804} & \multicolumn{1}{c|}{+2184} & +1539 & \multicolumn{1}{c|}{792} & \multicolumn{1}{c|}{-515} & -355 & \multicolumn{1}{c|}{$\sqrt{}$} & \multicolumn{1}{c|}{$\sqrt{}$} & \multicolumn{1}{c|}{$\sqrt{}$} & $\sqrt{}$ & \multicolumn{1}{c|}{1108} & 10/10 \\ \cline{1-1} \cline{3-17} 
\multicolumn{1}{|c|}{13} & \multicolumn{1}{c|}{} & \multicolumn{1}{c|}{Tip} & \multicolumn{1}{c|}{R} & 12*8 & \multicolumn{1}{c|}{854} & \multicolumn{1}{c|}{+630} & +472 & \multicolumn{1}{c|}{892} & \multicolumn{1}{c|}{-452} & -258 & \multicolumn{1}{c|}{$\sqrt{}$} & \multicolumn{1}{c|}{$\sqrt{}$} & \multicolumn{1}{c|}{$\sqrt{}$} & $\sqrt{}$ & \multicolumn{1}{c|}{1108} & 10/10 \\ \cline{1-1} \cline{3-17} 
\multicolumn{1}{|c|}{14} & \multicolumn{1}{c|}{} & \multicolumn{1}{c|}{Pulp} & \multicolumn{1}{c|}{R} & 10*8 & \multicolumn{1}{c|}{818} & \multicolumn{1}{c|}{+463} & +393 & \multicolumn{1}{c|}{936} & \multicolumn{1}{c|}{-230} & -197 & \multicolumn{1}{c|}{$\sqrt{}$} & \multicolumn{1}{c|}{$\sqrt{}$} & \multicolumn{1}{c|}{$\sqrt{}$} & $\sqrt{}$ & \multicolumn{1}{c|}{1284} & 8/10 \\ \cline{1-1} \cline{3-17} 
\multicolumn{1}{|c|}{15} & \multicolumn{1}{c|}{} & \multicolumn{1}{c|}{Palm} & \multicolumn{1}{c|}{R} & 8*14 & \multicolumn{1}{c|}{1400} & \multicolumn{1}{c|}{+3011} & +2130 & \multicolumn{1}{c|}{1240} & \multicolumn{1}{c|}{-2153} & -1859 & \multicolumn{1}{c|}{$\sqrt{}$} & \multicolumn{1}{c|}{$\sqrt{}$} & \multicolumn{1}{c|}{$\sqrt{}$} & $\sqrt{}$ & \multicolumn{1}{c|}{1443} & 10/10 \\ \hline
\end{tabular}
}

\begin{tablenotes}
    \footnotesize
    \item \textbf{Avg.:} Average interference weight across all sensing units (total weight derived by multiplying Avg. by the unit count); \textbf{Max.:} Maximum interference weight observed; 
    
    \textbf{R:} Resistive; \textbf{C:} Capacitive. \textbf{$\times$:} Indicates resilience to DoS attacks; \textbf{N/A:} Not applicable (single-unit sensors requiring no spatial control).
\end{tablenotes}

\end{table*}

\subsection{Overall Performance}

We evaluate our proposed \xxx attack on 15 COTS tactile sensors from 8 manufacturers, including 10 commercial sensor modules and 5 distinct sensors from 2 dexterous hands. Sensor information is summarized in \cref{tab-perform}.
Following common IEMI practice~\cite{wang2022ghosttouch,tu2019trick,zhang2024virtual,yang2024rethink}, we identified effective signal parameters via automated frequency sweeping with manual refinement. Specifically, for each target sensor with its original package, we swept frequencies from 700~MHz to 1.6~GHz in 1~MHz steps with a dwell time of 500~ms to evaluate both manipulation-induced measurement deviation and DoS attacks. To avoid excessive EM exposure and prolonged high-power output, we started at $-20$~dBm and increased the power in 5~dBm steps until a measurable deviation appeared. We then repeatedly tested frequencies within the susceptible range to identify the one producing the largest and most stable effect. For each sensor, we recorded the most effective frequency, the resulting measurement deviation, DoS success rate and the manipulation types. The results are summarized in~\cref{tab-perform}, where the checkmarks in the Force column denote cases in which the EMI-induced deviation exceeds the measurement noise.
The Isense UM1208 and Linker L10 Tip sensors were evaluated at a power level of 0 dBm, whereas all remaining sensors were tested at -10 dBm.
\begin{figure}[t]
    \centering
    \subfigure[Manipulation efficacy.]{
        \begin{minipage}[t]{0.45\linewidth}
            \centering
            \includegraphics[width=1\textwidth]{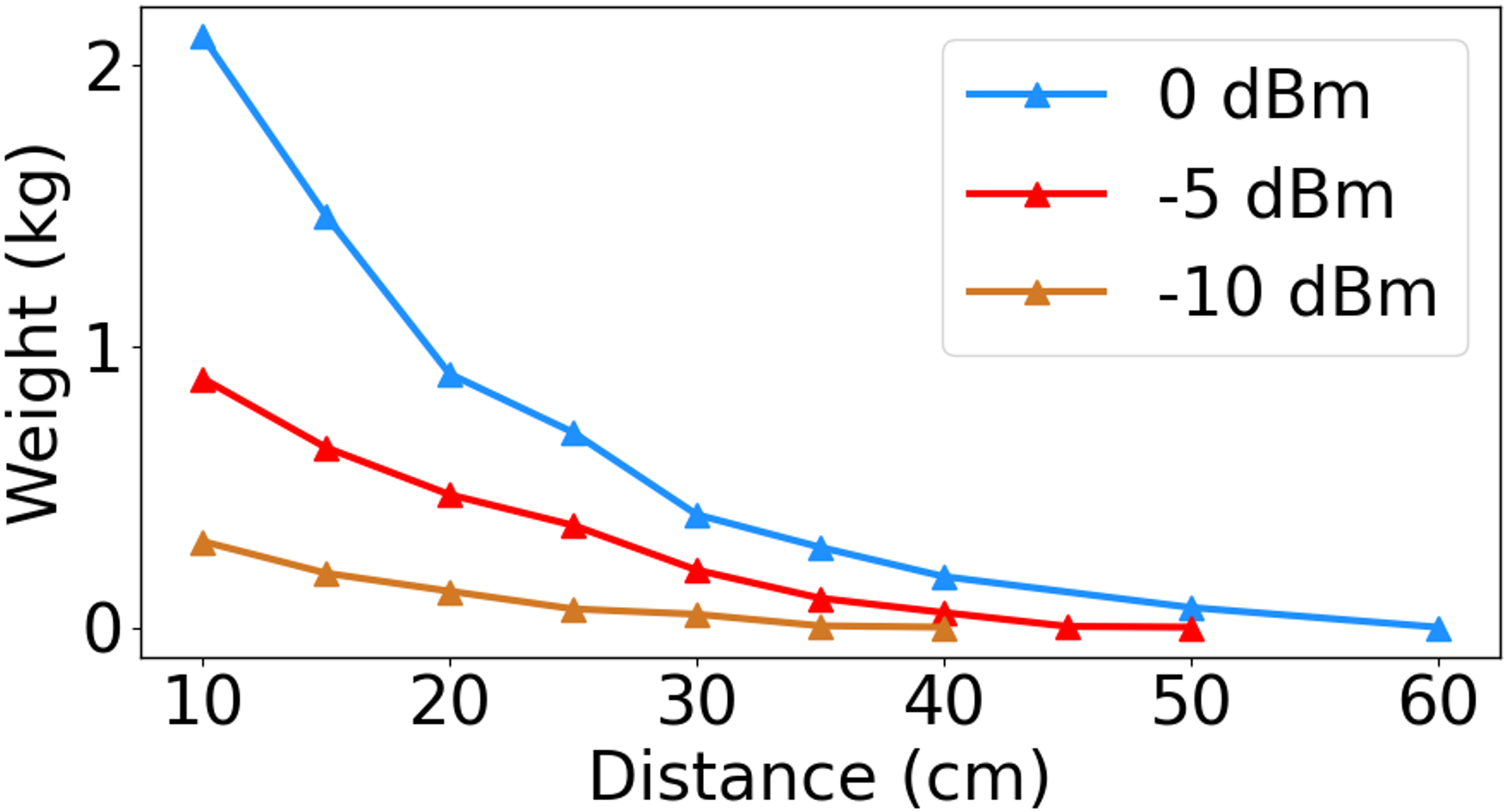}
        \end{minipage}
        \label{fig-distance1}
        
    }
    \subfigure[DoS efficacy.]{
        \begin{minipage}[t]{0.45\linewidth}
            \centering
            \includegraphics[width=1\textwidth]{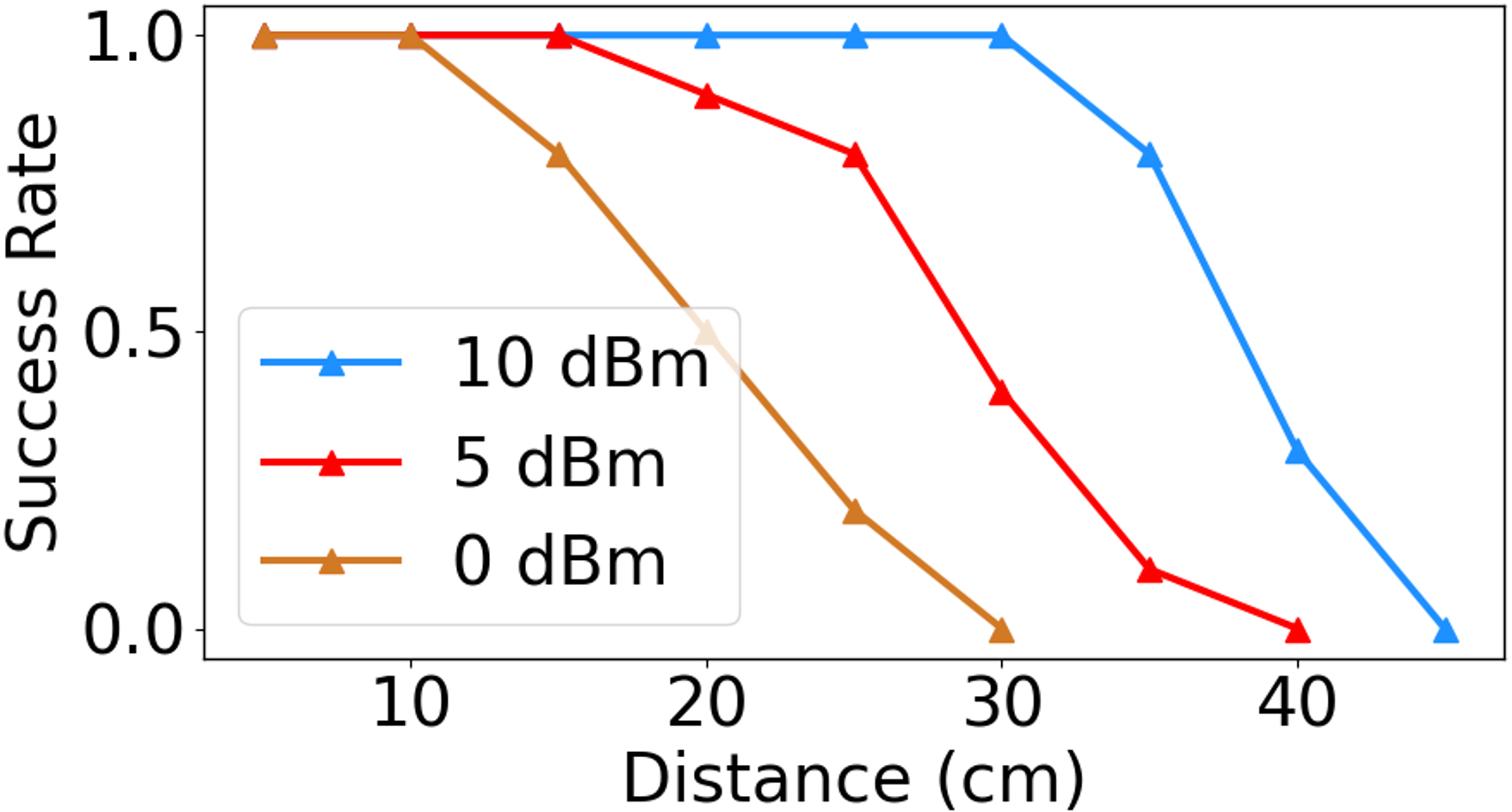}
        \end{minipage}
        \label{fig-distance2}
        
    }
    \label{fig-distance}
    \caption{Distance impact on manipulation and DoS. }
\end{figure}
The results demonstrate that all tested sensors could be successfully manipulated to exhibit positive or negative measurement deviations, and 13 out of 15 were susceptible to DoS attacks.
For example, when injecting a 1156 MHz signal into the RPPS-36 array, the most affected sensing unit reported a spurious weight of 4870 g, with a mean induced weight of 4812 g across all units. Given the array's 36 sensing units, this results in a total aggregate weight of approximately 173.2 kg. This magnitude is physically impossible to generate through normal interaction, but can be achieved by our proposed attack. Additionally, a 1325 MHz EMI signal triggered a DoS state, rendering the sensor unavailable.
Furthermore, we experimentally validated the four manipulation types proposed in \cref{sec-design-effects}. We successfully achieved static and dynamic effects on all tested sensor arrays, excluding single-unit sensors. This exclusion is because fine-grained spatial manipulations, other than force manipulation, are inherently inapplicable to individual units.

\subsection{Factors Affecting \xxx Attack}

We evaluated the impact of attack distance and attack angle on the performance of the \xxx attack in the palm sensor of a commercial dexterous hand~\cite{hand}.

\subsubsection{Attack Distance}
We evaluate the effectiveness of the force measurement manipulation and DoS attacks under different distances and injection power of the signal generator with a compact log-periodic antenna.

\textbf{Manipulation evaluation.} With the attack angle held constant, we varied the distance between the antenna and the sensor, starting at 10 cm in 5 cm increments, and repeated the measurement at three different power levels, as shown in~\cref{fig-distance1}. Measurements were continued until the observed deviation in reported weight returned to baseline, defined as the measurement noise range. Across all power levels, the magnitude of the induced deviation decreased monotonically with distance, consistent with expected electromagnetic attenuation during propagation.

\textbf{DoS evaluation.} For the DoS evaluation, we varied the distance starting from 5 cm in 5 cm increments. At each distance and power setting, we performed 10 DoS attempts to estimate the success rate. The results for different power settings are summarized in~\cref{fig-distance2}. The DoS success rate decreased with increasing distance and lower injection power, while at the highest power level, it remained achievable up to 40 cm.

Overall, these results indicate that both the ability to manipulate with measured weights and the probability of inducing a DoS decrease with distance, while higher injection power extends the effective attack range.

\subsubsection{Attack Angle}
\begin{figure} \label{fig-angles}
    \centering
        \subfigure[Vertical angle.]{
        \begin{minipage}[t]{0.45\linewidth}
            \centering
            \includegraphics[width=1\textwidth]{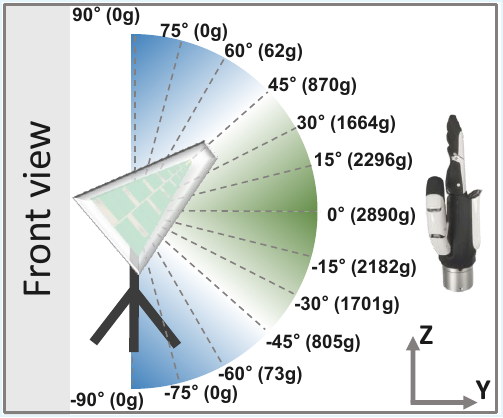}
        \end{minipage}
        \label{fig-angle1}
    }
    \subfigure[Horizontal angle.]{
        \begin{minipage}[t]{0.45\linewidth}
            \centering
            \includegraphics[width=1\textwidth]{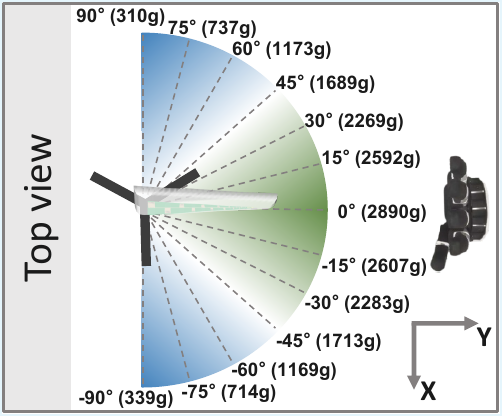}
        \end{minipage}
        \label{fig-angle2}   
    }
    \caption{The influence of attack angles.}
\end{figure}

Aiming precision is a common challenge for remote attacks; the lower the required precision, the more practical the attack becomes. We evaluated angular tolerance by sweeping the antenna from −90° to 90° in 15° increments at a distance of 10 cm, recording the induced deviation at each angle. Measurements were performed for both vertical (shown in~\cref{fig-angle1}) and horizontal (shown in~\cref{fig-angle2}) orientations, with a baseline reading of 0 g.

Our results show that force measurement manipulation was successful within a ±60° vertical angular window. For the horizontal orientation, the attack was even more robust; a deviation of 310 g was observed even at 90°. These findings indicate that the attack does not require precise aiming and remains feasible under realistic deployment conditions.

\subsubsection{Long-Range and Through-Obstacle Evaluation}
\begin{figure}[t]
    \centering
    \includegraphics[width=0.95\linewidth]{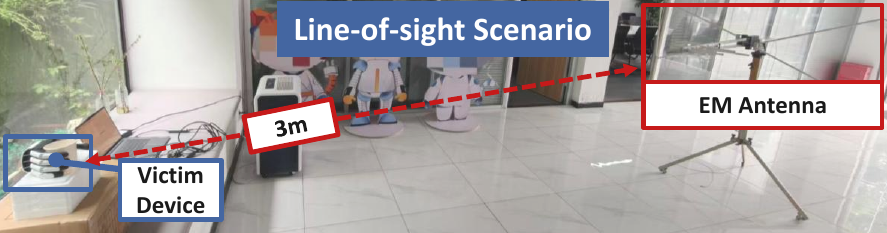}
    \caption{The long-range attack setup under LoS scenarios.}
    \label{fig-new_scenario2}
\end{figure}

To further explore the maximum attack range, we used dedicated hardware to evaluate long-range performance. Specifically, we conducted experiments with a high-gain polarized log-periodic antenna (VULB 9162~\cite{schwarzbeck-vulb9162}), as shown in~\cref{fig-new_scenario2}. Under line-of-sight conditions, GhostTac achieved bottle-squeezing attacks at distances up to 3 m, as shown in~\cref{fig-new_scenario2}. 
We quantified the achievable deviation as 426 g per sensing unit. The total deviation scales with the number of contact units (6–10 sensing units in this task); therefore, it is sufficient for bottle-squeezing.
To evaluate GhostTac under more realistic deployment conditions, where obstacles may block the direct line of sight, we tested it in several obstructed scenarios. GhostTac successfully enabled bottle squeezing through a window, a wooden door, and a wall.

\subsection{Case Studies}
To demonstrate the threat of \xxx in real tasks, we evaluate three representative case studies, including grasping, slip detection, and material classification. Each task employs a Franka Emika Panda arm~\cite{franka-robotics} mounted with a dexterous hand~\cite{hand}.
\subsubsection{Grasping}
\begin{figure}[t]
    \centering
    \includegraphics[width=1.0\linewidth]{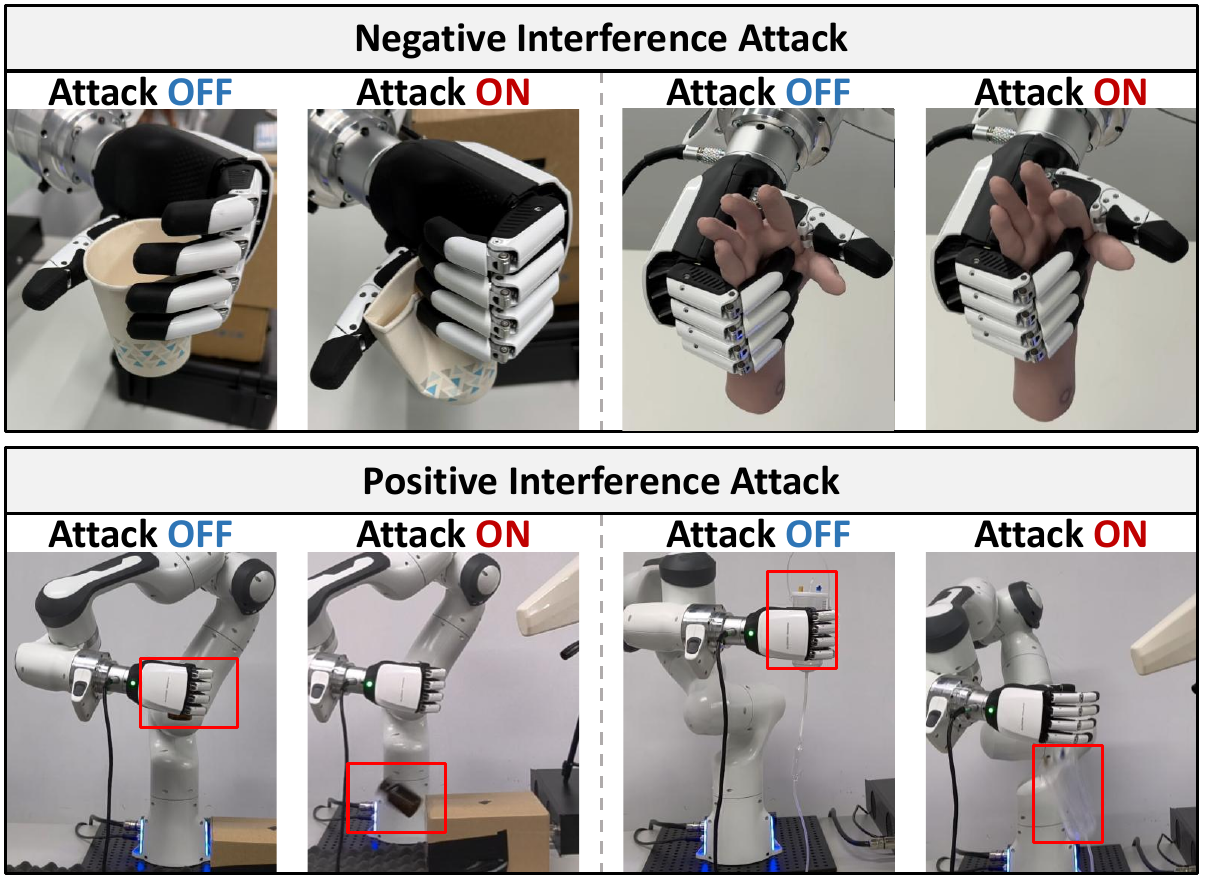}
    \caption{Illustration of the grasping conditions before and after the attack. Negative interference causes excessive force application, leading to object damage, while positive interference reduces grip force, causing the object to drop.}
    \label{fig-grasping}
\end{figure}

We implement a tactile feedback controller that maintains grasp stability by regulating the measured force within predefined thresholds and adapting to deviations.
Then, we apply this adaptive strategy in robotic arm grasping tasks, where attacks are conducted along the grasping path. We evaluated two attack modes where negative interference triggers excessive force application, leading to object damage, and positive interference reduces grip force, resulting in object drops. Comparisons between normal and attacked grasps are illustrated in~\cref{fig-grasping}. Under normal operation, objects were grasped stably without damage. Under attack, excessive force deformed the simulated silicone hand and a paper cup, while insufficient force caused the intravenous infusion device and glass bottles to drop.  Both attack types produced clear and consistent effects in all ten trials due to the broad EMI coverage that reduced the need for precise targeting. 
A successful attack is defined as either visible deformation of the grasped object or object release. In the plastic-bottle example, visible squeezing requires a total weight deviation of more than 2240 g (a 20 g average deviation per sensing unit), whereas release requires a deviation of more than 784 g (a 7 g average deviation per sensing unit).

\begin{figure}[t]
    \centering
        \subfigure[Slip detection.]{
        \begin{minipage}[t]{0.45\linewidth}
            \centering
            \includegraphics[width=1\textwidth]{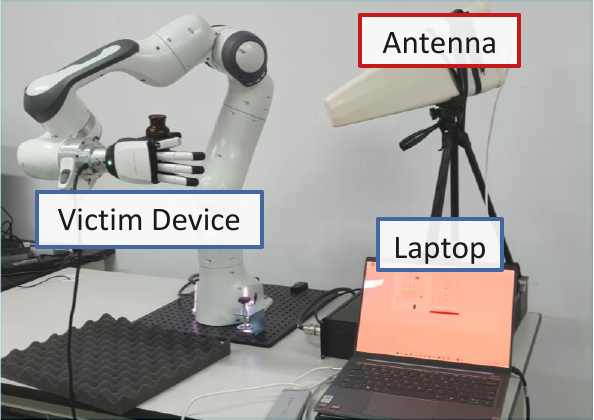}
        \end{minipage}
        \label{fig-setupcase1}
    }
    \subfigure[Material classification.]{
        \begin{minipage}[t]{0.45\linewidth}
            \centering
            \includegraphics[width=1\textwidth]{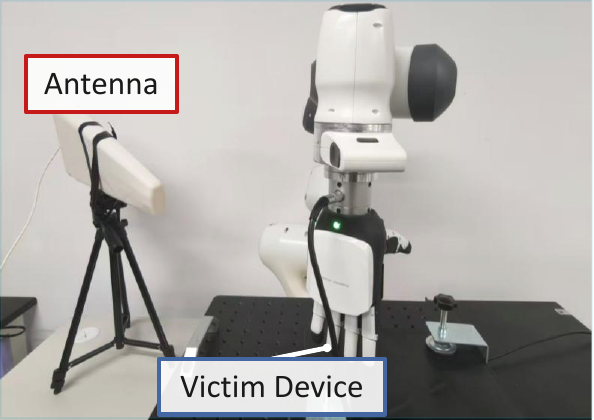}
        \end{minipage}
        \label{fig-setupcase2}   
    }
    \caption{The setup for case studies.}
\end{figure}

\subsubsection{Slip Detection}
The slip-detection method used here operates by tracking the displacement of the center of pressure (CoP) over time, which is computed from a tactile sensor array. 
This approach relies on the principle that incipient slip causes shifts in the pressure distribution at the contact interface. Our algorithm classifies the grasp state based on the direction and consistency of CoP displacement. Persistent shifts indicate sliding, whereas a stable CoP indicates a secure grasp.

\begin{figure}[t]
    \centering
    \includegraphics[width=1.0\linewidth]{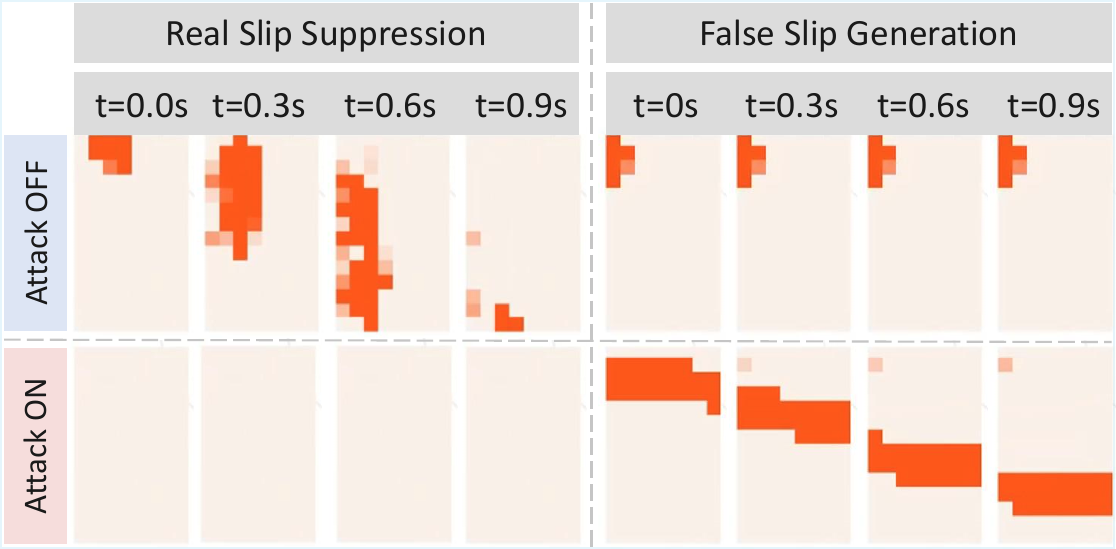}
    \caption{The injected weight distributions under slipping attack, illustrating real slip suppression and false slip generation effects on the sensor.}
    \label{fig-slipping}
\end{figure}

We evaluated \xxx on a robotic hand gripping a medical bottle as shown in~\cref{fig-setupcase1}. We first validated the algorithm's accuracy through 10 stable grasping events and 10 sliding events, all of which were correctly classified. Our validation demonstrated two distinct, successful attack scenarios (shown in~\cref{fig-slipping}). A successful attack is defined as causing either a false positive, where a slip is detected when none occurs, or a false negative, where a real slip is not detected.
In the scenario regarding real slip suppression, the system output was manipulated to indicate a stable grip state while the object was slipping.  
Conversely, in the false slip generation scenario, the system was driven from a no-slip state into a slip classification by the injected attack pattern, which the detector interpreted as actual slipping. Each test was repeated 10 times with consistent results.

\subsubsection{Material Classification}
We implement a Long Short-Term Memory (LSTM) model to classify materials from time-series data recorded by a 3×3 tactile array during constant velocity sliding. The dataset comprises 80 samples over four materials, which we partitioned for training (80\%) and validation (20\%).  Raw nine-channel pressure traces are windowed into 50-time-step segments and fed to the LSTM, which ends in a Softmax layer that outputs class probabilities. For interference on an unknown material, a new 50-time-step window is classified based on the model's highest probability output. 

\begin{figure}[t]
    \centering
    \includegraphics[width=1.0\linewidth]{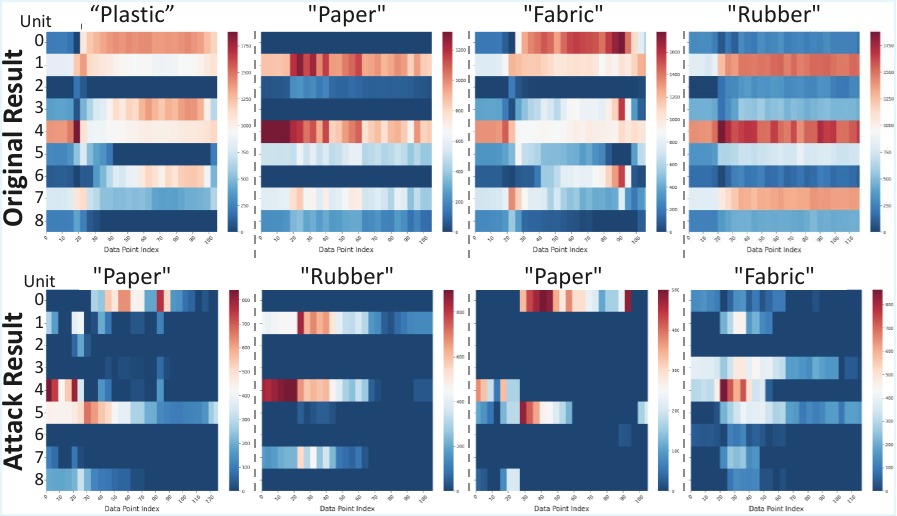}
    \caption{The weight variation over time of nine tactile sensing units and classification results. 
    Each column represents a specific material, while the top and bottom rows show original and attack results, respectively.}
    \label{fig-casestudy_material}
\end{figure}

We conduct a sliding test where the sensor scans the target material under signal injection, as shown in~\cref{fig-setupcase2}.
Following prior attack methods, experiments on four materials demonstrate successful manipulation of recognition results across all cases, and detailed comparisons between the original and attacked results for all nine sensing units are provided in~\cref{fig-casestudy_material}.

\subsection{Real-world Attack Scenarios}
We present two real-world attack scenarios to highlight the practicality of our proposed method. In the first scenario, an attacker approaches a target device, such as a medical robot in a hospital, while concealing the attack hardware in a bag and emitting the signal while passing. 
The transient and concealed execution improves covertness and minimal traceability. If the attacker cannot approach the victim's device during operation, they can conceal the device nearby, for example, by attaching it under a table, and trigger it remotely to disrupt grasping tasks. This remote activation may improve stealth and ensure the practicality of the attack. Both scenarios demonstrate the feasibility and potential impact of the attack in realistic environments.
Furthermore, a more compact implementation could improve attack portability and stealth.

%% file: sections/discussion.tex
\section{Discussion}

\subsection{Countermeasures}
Potential countermeasures against \xxx can be categorized into two directions, i.e., strengthening the electromagnetic robustness and enhancing the resilience of tactile sensing mechanisms.

\subsubsection{Enhance the EMC}

Three common strategies are employed to enhance the EMC of tactile sensors: shielding, differential comparators, and filtering. Shielding encloses components in conductive materials to suppress EMI based on the Faraday cage principle~\cite{chapman2015mathematics}. 
While shielding is effective, it is frequently impractical and insufficient. Differential circuits address this by leveraging a complementary signal to cancel the common-mode voltage induced by EMI.
Filters suppress frequencies outside the baseband to reduce interference, but conventional first-order low-pass filters often fail against high-frequency signals due to parasitic capacitance. In such cases, radio-frequency filters are needed for stronger suppression~\cite{analog2009rfi}.  However, these methods necessitate additional hardware and consequently increase system complexity and cost.

\subsubsection{Enhance the Sensing Mechanism.}

Three defenses can be employed to improve sensing robustness against EM attacks:
(1) Increasing excitation signal voltage improves the signal-to-noise ratio, thereby reducing the impact of EMI attacks and making it more difficult for attackers to inject malicious signals capable of disrupting sensor measurements.
(2) Randomizing the excitation waveform mitigates vulnerabilities arising from the processor’s inability to authenticate the received signal source.
(3) Randomizing scan timing $T_p$ makes the process unpredictable, preventing attackers from determining the scan moment or electrode and thus reducing the feasibility of precise injections and targeted attacks.

\subsection{Limitations and Future Work}

While \xxx shows promising results, there remain several limitations for further investigation.

\subsubsection{Spatial Control} 
Because of the driving schemes and wide EM coupling between RXs, the \xxx attack can only produce controllable interference along TXs and cannot be reliably deployed at arbitrary locations.
Nevertheless, the current search range, both in location and intensity, is sufficient to construct effective adversarial attacks for diverse scenarios. Moreover, when coupling characteristics vary, such as in resonant frequencies or efficiencies specific to columns, an attacker can still reliably target RX.

\subsubsection{Sensor Modality Scope} Our experiments primarily focused on piezoresistive and capacitive tactile sensors, as these types dominate the current market~\cite{li2024electromechanics,xi2024recent}. Further investigation will be conducted to determine whether visuotactile and other multimodal sensors exhibit similar EM vulnerabilities.

\subsubsection{Angular Misalignment and Motion}
\xxx remains effective within a spatial region around the target sensor rather than only at a single precisely aligned position. During moving-grasping tasks, the attack can still induce observable weight deviations as long as the target sensor remains within a susceptible spatial region. Nevertheless, the attack effect degrades as misalignment increases because the resulting misalignment weakens the EM coupling between the transmitting antenna and the target sensing circuitry.
In highly dynamic scenarios with rapid movement, the attack effect is expected to diminish as stable EM coupling becomes harder to maintain. Supporting such settings would likely require tracking-assisted targeting~\cite{sato2025realism}, where the attacker uses an external vision-based tracker to estimate target motion and a servo-controlled steering system to continuously adjust antenna orientation and maintain effective coupling.

\subsection{System-Level Analysis}
This paper focuses on exposing a modality-level vulnerability in tactile sensing. Future robotic systems may be resilient to attack due to system-level mechanisms such as sensor fusion, closed-loop control, and anomaly detection. We therefore discuss the adaptability and limitations of \xxx in the presence of each of these system-level designs.
\subsubsection{Sensor Fusion}
Sensor fusion combines tactile sensing with other modalities, such as vision and proprioception, to improve perception robustness and task performance. In principle, such cross-modal redundancy may reduce the effectiveness of \xxx, since inconsistent tactile measurements may be mitigated by information from other sensors.
However, the protection offered by sensor fusion is inherently task-dependent. Prior work has shown that some tasks, such as material recognition~\cite{kerr2018material} or manipulation in camera-blind regions, rely primarily on tactile feedback rather than vision. In these cases, system decisions are dominated by tactile measurements and therefore remain vulnerable to \xxx. Moreover, an adversary could potentially combine \xxx with existing attacks on other sensing modalities, such as camera blinding~\cite{yan2022rolling}, to force the system to rely on compromised tactile input.
\subsubsection{Closed-loop Control}
Closed-loop control refers to systems that continuously adjust actions according to tactile feedback. If tactile measurements are directly trusted for force adjustment, contact regulation, or control-state transitions, \xxx can manipulate the loop in real time by injecting continuous forged tactile signals. For the grasping tasks in our case studies, we further demonstrate that \xxx can sustain such manipulation over consecutive control cycles and continuously mislead the closed-loop control algorithm. Defending against this threat remains challenging, since simple filtering is insufficient against carefully crafted continuous injections. More reliable mitigation likely requires stronger safeguards around tactile-driven control, such as trusted contact verification, active physical consistency probing, or runtime monitoring of suspicious temporal patterns.
\subsubsection{Anomaly Detection}
Anomaly detection refers to runtime mechanisms that identify tactile signals deviating from expected measurements, e.g., based on abnormal magnitudes, temporal inconsistency, or spatial patterns. In principle, such mechanisms could mitigate \xxx. However, simple detection based on signal amplitude or temporal continuity is unlikely to be reliable, because the attacker could regulate the injected force in real time and keep forged tactile readings within plausible ranges. A more promising direction is pattern-based detection. Specifically, GhostTac can only generate controllable interference along TXs and cannot be reliably placed at arbitrary locations. Defenders may therefore distinguish attacks from genuine contacts by checking whether the observed activation pattern is physically consistent with natural contact geometry.

%% file: sections/related_work.tex
\section{Related Work} \label{related_work}

Sensors are vulnerable to EMI attack, and it has been widely studied in the security research community to destroy the integrity and reliability of sensor outputs in recent years. For manipulating the sensor’s measurement,  EMI attacks have been reported on microphone~\cite{kune2013ghost,dai2023inducing,xiao2025sok}, touchscreen~\cite{jiang2022wight,wang2022ghosttouch,shan2022invisible}, temperature sensor~\cite{tu2019trick,lavau2023securing}, Lidar~\cite{jin2024phantomlidar,bhupathiraju2023emi}, image sensor~\cite{dai2023magcode,jiang2023glitchhiker}, and so on. The consequences of these attacks range from denial-of-service to injecting malicious data or even completely manipulating the operations of sensor-based cyber-physical systems. 
Apart from direct sensor attacks, various EMI attacks have been shown to break communication systems~\cite{liu2024nfceraser,jang2023paralyzing,kohler2022brokenwire}. These attacks can also manipulate PWM driving signals in converters and servo motors~\cite{dayanikli2020electromagnetic,dayanikli2022physical}. Additionally, some approaches exploit other components to transmit EM signals indirectly, thereby affecting sensor operations~\cite{dai2023inducing,zhan2024voltschemer,yang2025lightantenna}.

\textbf{Novelty over prior work.} Our work focuses on uncovering a new attack vector against tactile sensing in emerging robotic systems. Despite the existing IEMI studies, exploiting tactile sensors is non-trivial due to their unique characteristics. Unlike the threshold-based sensing of touchscreens~\cite{wang2022ghosttouch,shan2022invisible}, tactile sensors measure minute force variations at dense locations, which are processed by various algorithms to support complex robotic tasks. To overcome these challenges, \xxx introduces a parameterized modulation framework that enables fine-grained manipulation of force magnitude, polarity, location, and sensor selection, with most of these capabilities achieved for the first time on grid-style sensors. We also design dynamic adversarial patterns to compromise system-level robotic functionalities, e.g., slip detection and material classification. In addition, we contribute to the IEMI mechanism with circuit-level analysis and physical experimental validation of nonlinear rectification and limited-bandwidth amplification effects, which were unmentioned or only briefly discussed in prior work~\cite{shan2022invisible,tu2019trick,zhang2024virtual,yang2024rethink}.

%% file: sections/conclusion.tex
\section{Conclusion}

In this paper, we present \xxx, the first contactless attack that manipulates tactile sensor measurements via EMI. This framework precisely controls the force magnitude and location, or induces DoS regardless of whether the target sensor is physically touched or moved.
We analyze the underlying principle theoretically and experimentally. The results indicate that coupling vulnerabilities in the sensing array combine with nonlinear rectification to convert the EMI signal at high frequencies into a DC offset that corrupts force measurements. 
We validate the effectiveness of \xxx on 10 COTS sensor modules and two dexterous hands. Additionally, we propose practical countermeasures based on hardware and software designs that significantly reduce the risk.